\documentclass[apj]{emulateapj}
\usepackage{natbib}
\usepackage{mathptmx}
\usepackage{url}
\usepackage{graphicx}	
\usepackage{amsmath}	
\usepackage{amssymb}
\usepackage[toc,page,titletoc]{appendix}
\usepackage[dvipdfm]{hyperref}
\begin{document}
\nocite{*}

\title{Evolution of Galaxy Types and HI Gas Contents in Galaxy Groups}
\author{Mei Ai$^{1, 2,3}$,Ming Zhu$^{1,2*}$ }
\affil{$^{1}$National Astronomical Observatories, Chinese Academy of Sciences ,20A Datun Road, Chaoyang District, Beijing, China}
\affil{$^{2}$CAS Key Laboratory of FAST, NAOC, Chinese Academy of Sciences}
\affil{$^{3}$University of Chinese Academy of Sciences, Beijing 1000101,China}

\email{$^{*}$mz@nao.cas.cn}

\date{Accepted XXX. Received YYY; in original form ZZZ}

\label{firstpage}

\begin{abstract}

Using the group crossing time $t_{\rm c}$ as an age indicator for galaxy groups,
we have investigated the correlation between $t_{\rm c}$ and the group spiral fraction, as well as between $t_{\rm c}$ and the neutral hydrogen gas fraction of galaxy groups.  Our galaxy group sample is selected from the SDSS DR7 catalog,  and the group spiral fraction is derived from the Galaxy Zoo morphological data set. We found that the group spiral galaxy fraction is correlated with the group crossing time. We further cross matched the latest released ALFALFA 70\% HI source catalog with the SDSS group catalog and have identified 172 groups from the SDSS survey whose total HI mass can be derived by summing up the HI mass of all the HI sources within the group radius. For the  galaxies not detected in the ALFALFA, we estimate their HI masses based on  the galaxies' optical colors and magnitudes. Our sample groups contain more than 8 member galaxies, they cover a wide range of halo masses and are distributed in different cosmic environments. We derived the group HI mass fraction which is the ratio of group HI mass to the group virial mass. We found a correlation between the HI mass fraction and the group crossing time. Our results suggest that long time scale mechanisms such as starvation seem to play a more important role than short time scale processes like stripping in depleting HI gas in the SDSS galaxy groups.

\end{abstract}



\section{Introduction}

In the hierarchical galaxy formation model,  disk galaxies form first and then fall into bigger halos. By merging, dark matter halos grow over cosmic time.  Galaxies tend to appear in  groups, or clusters. Groups have fewer galaxy members than clusters, but can grow into clusters when more members are accumulated.  Thus, they could be considered as an intermediate system in the Universe. These systems not only probe the Large Scale Structure of matter distribution, but also offer a unique environment for studying their impact on galaxy evolution. Modern large sky surveys such as the Sloan Digital Sky Survey (SDSS) \citep{York2000} have detected a large number of galaxies and thousands of galaxy groups have been identified  in several group catalogs (e.g. \citealt{Berlind2006}; \citealt{Crook2008}), which enable us to make a statistical analysis of the relationship between group properties and their environments.

Neutral hydrogen (HI) gas is an important component of galaxies. This component is loosely gravitationally bound, and is easily disturbed by galaxy interactions and tidal forces \citep*{Hibbard1996}. When galaxies fall into the gravitational potential center of a group or a cluster, HI gas may be heated or ionized by shocks, and may also escape the interacting system and disperse, becoming too diffuse to detect \citep*{Hibbard1996}.  Galaxy merging can also efficiently remove angular momentum so that HI clouds can be accreted into the central part of galaxies, increase their density and cool quickly to form molecular clouds. Therefore, HI content is a good tracer of a group's internal interaction level, and should be related to the evolutionary stages of galaxy groups or clusters.

Many investigators have noticed that galaxy groups and clusters are deficient in HI (\citealt{Verdes-Montenegro2001},\citealt{Solanes2004},\citealt{Taylor2012}). Many previous works focus on the relationship between gas content and group/cluster environment(\citealt{Davies1973},\citealt{Haynes1984},\citealt{Solanes2001},\citealt{Kilborn2009},\citealt{Rasmussen2012},\citealt{Serra2012},\citealt{Brown2016},\citealt{Stark2016}). Quantitatively, the HI deficiency is defined as a measure of how much gas a galaxy of a given morphological type and optical diameter has lost in comparison to a similar field galaxy (\citealt{Haynes1984}; \citealt{Giovanelli1985}). Several mechanisms have been suggested to explain the environmental influence on HI deficiency such as ram pressure stripping (\citealt{Gunn1972},\citealt{Kenney2004}, \citealt{Boselli2006}, \citealt{Chung2009}, \citealt{Cortese2011}), galaxy harassment \citep{Moore1996} and interactions between individual galaxies \citep{Mihos2004}, starvation (\citealt{Larson1980},\citealt*{Balogh2000},\citealt*{Bekki2002},\citealt*{Kawata2008}) and viscous stripping (\citealt{Rasmussen2012}), but their relative importance is still not well understood. The detailed physical processes for group HI deficiency remain unclear (\citealt{Catinella2013},\citealt{Hess2013}). After analyzing the HI content of spirals in 18 nearby clusters,  \cite{Solanes2004} concluded that the observed distribution of HI deficiency in the Virgo core appears to be closely associated with the dynamically evolutionary state of the main aggregates.

Another observational fact is that galaxy properties such as optical colors, morphologies and star formation rates are closely correlated with the galaxy environment when measured by the galaxy number density (\citealt{Lewis2002}, \citealt{Kauffmann2004},  \citealt{Blanton2009}, \citealt{Bretherton2013}, \citealt{Pasetto2014}, \citealt{Mok2016}). For example, \cite{Dressler1980} pointed out that the fraction of early type and S0 galaxies increases with increasing environmental density. \cite{Balogh1997} also found that the fraction of star forming galaxies is smaller in cluster environment than that in the field.

Most galaxy evolution studies are based on observations of galaxies at different redshifts, such as the Butcher-Oemler effect (\citealt{Butcher1978}; \citealt{Butcher1984}), which shows an increase in the blue cluster population with redshift.
Observations of molecular gas traced by CO have shown that as redshift goes higher there is a statistical increase of CO gas fraction in galaxies (c.f. the IRAM Plateau de Bure high-z blue sequence CO 3-2 survey, \cite{Tacconi2013} and references therein).
Many investigators have attempted to detect HI emission at high redshift.
However, because of the inherent weakness of the HI line and the contamination of the spectrum by terrestrial radio frequency interference (RFI), studies of redshifted HI emission from individual galaxies have only recently been possible up to 0.25-0.37 (\citealt{Catinella2013}; \citealt{Fernandez2016}). Beyond that, the study of HI emission from individual objects requires the resolution and collecting area of the future Square Kilometer Array (SKA).

As pointed out by \cite*{Ai2017} the sensitivity of large single dish telescopes, such as the  Five-hundred-meter Aperture Spherical radio Telescope(FAST, \citealt{Nan2011}) is high enough to detect  HI emission from galaxy groups and clusters at intermediate redshift. At redshift of 0.7, the beam size of FAST is $2.95'(1+z)$ corresponding to a physical size of about 0.9 Mpc which is the typical scale of a group or cluster, thus FAST can be used to measure the integrated HI emissions from groups and to study the evolution of the group total HI contents over cosmic time.

In order to establish a benchmark for studying high redshift groups, we have carried out a systematic study of the global HI properties in a large sample of low redshift galaxy  groups derived from the SDSS surveys.
Our study shows that galaxy evolution in group environment can also be studied in the local universe. For example, \cite{Hickson1982} published a catalog of hyper compact galaxy groups, and they found that the fraction of spiral galaxies is well correlated with their evolutionary status characterized by the quantity of group crossing time \citep{Hickson1992}. This suggests that the "Group crossing time" $t_{\rm c}$ could be used as a criterion to distinguish group dynamical states. Since spiral galaxies are normally gas rich,  we would expect that the HI gas fraction in groups is also correlated with the group crossing time. In this paper we explore the correlation between $t_{\rm c}$ and the group spiral fraction, as well as between $t_{\rm c}$ and the neutral hydrogen gas fraction of galaxy groups.


In all the calculations we use the Hubble constant $H_0$ = 70 $\rm km\;s^{-1}Mpc^{-1}$.

\section{The group sample and related data }
\subsection{Galaxy Group Catalogs}

Our sample was selected from the SDSS DR7 Mr18 group catalog. This catalog is based on an optimized algorithm of the friends of friends (FoF) method which is described in detail in \cite{Berlind2006}. The complete group catalogs derived from SDSS DR7 are available online (\url{http://lss.phy.vanderbilt.edu/groups/dr7/}). There are three galaxy samples created with different absolute magnitude limits and redshift ranges. Each one is complete within the stated limits. Given that the highest redshift of the HI sources in the ALFALFA \citep{Giovanelli2005} is around 0.06 and the luminous HI source corresponds to less luminous optical galaxy, we choose the faintest Mr18 group sample whose r band absolute magnitude is down to $M_r=-18$. The redshift range for the SDSS Mr18 group is between 0.02 and 0.042. We restrict our sample to groups with galaxy member count N$\ge$ 8 because the virial mass of very small group is extremely unreliable due to large uncertainties in determining the group radius and velocity dispersion. We delete the "groups" which have N$>$200 because their properties are similar to big clusters, and thus we did not include them in our group sample. With these criteria we selected 7367 galaxies within 459 groups.

\subsection{Morphology of group member galaxies}
The morphological data come from the Galaxy Zoo 1 \citep{Lintott2011} data set. There are nearly 700,000 SDSS DR7 classified galaxies with available spectra and the results are debiased based on the redshift information. Over 200,000 DR7 classified galaxies have no spectra, thus no accurate redshift and  no bias estimate can be obtained. The huge number of classified galaxies makes Galaxy Zoo 1 the largest data set of galaxy morphologies and allows us to classify the morphology of most of our SDSS group member galaxies. The cross match of the SDSS member galaxies and Galaxy Zoo 1 galaxies was performed using the software TOPCAT with an angular distance of 5$''$. A total of 6552 SDSS member galaxies  in 440 groups with number of member galaxies $N\ge8$ are classified as spirals or ellipticals. We used a cut off value of 0.5 as morphological criterion to distinguish between the elliptical and spiral types in the Galaxy Zoo 1 data set.  Among the 6552 galaxies, 3357 galaxies are classified as spirals and 2819 galaxies are classified as ellipticals. The remaining 376 galaxies do not meet the spiral or elliptical criterion and we marked them as uncertain. We exclude those galaxies marked as "no-spectra" in the Galaxy Zoo 1 catalog in our sample because no debias was carried out for them. So, a total of 95.86\% of our SDSS groups and 88.94\% of group members are identified with galaxy morphology types. According to \cite{Lintott2011}, a cut off value of 0.5 yields a misclassification rate of 19\%.  We also tried to use a cut off value of 0.6 as a morphological criterion which yields a misclassification rate of 10\%. In this case, 3130 galaxies are classified as spirals and 2655 galaxies are classified as ellipticals. The rest of 767 galaxies remain unclassified, representing 13\% of the sample galaxies.

\subsection{HI mass in SDSS groups}

\subsubsection{HI sources associated with SDSS groups}

To get the group HI mass, we use the recently released ALFALFA 70 ($\alpha.70$) catalog (for details see the reference to the $\alpha.40$ catalog by \cite{Haynes2011}), which contains 70\% of  the data from the ALFA extragalactic HI survey using the Arecibo 305 meter telescope. The $\alpha.70$ HI source catalog is available on the web site \url{http://egg.astro.cornell.edu/alfalfa/data/}. In this catalog, the detected HI sources with the available SDSS optical counterparts (OCs) have been identified and their (RA, DEC) coordinates are listed.  Similar to the process of \cite{Hess2013}, we assigned HI sources to group dark matter halos in a two step process.

In the first step we cross match the Mr18 group member galaxies with the $\alpha.70$ HI sources using TOPCAT, and the angular distance between SDSS member galaxies and $\alpha.70$ HI's OC is 5$''$.  We delete the matched pairs which have redshift difference larger than $\rm 200\;km\;s^{-1}$. There are 256 N$\ge$ 8 SDSS groups that lie in the $\alpha.70$ survey region and we matched 726 $\alpha.70$ HI sources with SDSS N$\ge$ 8 group member galaxies which are distributed in 216 SDSS groups.

In the second step we add several HI sources which are luminous in the radio band while faint in the optical band. These objects are not included in our group sample because their r band magnitudes are over the magnitude limit of the SDSS Mr 18 sample. We associate the $\alpha.70$ objects to our sample groups based on the following criteria: (1) have  HI detections where no OC is present in the $\alpha.70$ catalog or the magnitudes of the OCs are below our sample magnitude limit; (2) the projected distance between the $\alpha.70$ object and the SDSS group center is less than 2$R_{\rm rms}$, where $R_{\rm rms}$ is the rms radius defined in \citealt{Berlind2006}; (3) the velocity difference between the HI source and the group is less than the max value of the group velocity dispersion and $\rm 300\;km\;s^{-1}$. In this way we obtained 40 extra HI members in the previously identified 216 groups, adding  5\% in the total number of galaxies. The method used here for assigning HI sources to groups is very similar to that of \cite{Hess2013}.

For the SDSS-HI matched groups, we include both  ALFALFA "code 1"  and  "code 2" sources, which have a signal-to-noise ratio S/N $>$ 6.5, and
4.5 $<$ S/N $<$ 6.5, respectively.  They are likely to
be real because nearly all of them have a known optical counterpart at the
redshift of the HI source \citep{Haynes2011}.  The total HI mass of each group is the sum of all the HI detected galaxies in the group.

In Fig. 1 we compare the resulting HI group masses obtained with these two different methods, where $\rm Group\;M_{HI},1$ represents the HI mass obtained with the method in the first step described earlier in this section, and $\rm Group\;M_{HI},2$ represents the HI mass obtained with the two-step method. From this plot we can see that the difference in the group HI mass obtained by two different methods is small.

\subsubsection{HI source confusion and group edge effects}

It is reasonable to expect that confusion within the Arecibo beam would be a problem in group environments where galaxies are in close proximity. Table 2 of the $\alpha.40$ catalog \citep{Haynes2011} includes notes on sources for which the HI parameters are uncertain because of confusion or other issues. The ALFALFA pointing errors are on average $18''$ which is a fraction of the $3.5 \times 3.8'$ beam.

Of all the 766 HI sources in the 216 SDSS group members, 481 are from the $\alpha.40$ catalog and 51 of them are cataloged as "blended".  We remove 6 of these groups because they suffer from severe confusion effect and hence they are very difficult to deblend. Another 4 groups, in which only a pair of members are blended, are in close proximity and have similar HI flux. We choose to keep them in our sample and then use the larger HI member mass as the pair HI mass. There are 286 new HI sources from the $\alpha.70$, for which no confusion has been identified.  If the confusion rate is similar to  that of  the $\alpha.40$, we could estimate that about  3.7\%  HI members suffer from confusion. Therefore source confusion should have limited effects on our statistical results.

Another effect that needs to be considered is the edge effect which involves groups with optical members not fully included in the $\alpha.70$ survey field. We tried to identify such groups and remove them from our sample. For all groups, we exclude those that have more than 25\% members lying outside the $\alpha.70$ footprint. We restrict our sample to group virial mass greater than $10^{13}M_{\odot}$ which will be explained in detail in section 3.2. Ultimately after removing the groups with edge and confusion effects that have been noted in the $\alpha.40$, our SDSS Mr18 sample has 172 groups of 8 or more members which contains 730 HI galaxies.

\begin{figure}
\centering
    \includegraphics[height=3.0in,width=3.2in]{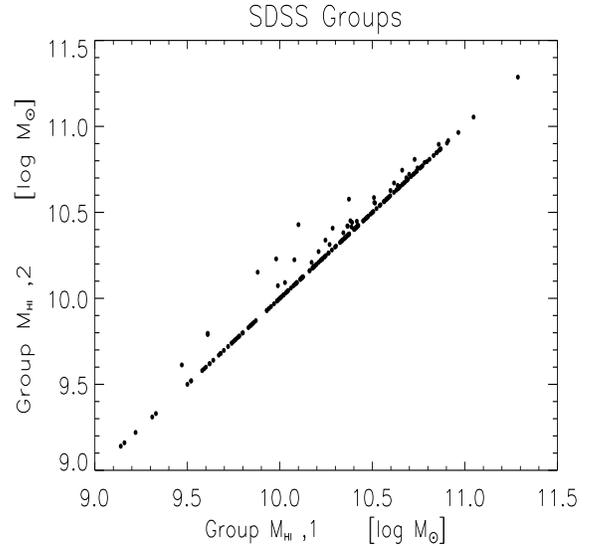}
    \caption{Comparison of the group HI mass obtained by two different methods: $\rm Group\;M_{HI},1$ refers to the one-to-one match method and $\rm Group\;M_{HI}.2$ refers to the mixed methods of the first and second steps in section 2.3.1 (see text for details).}
\end{figure}

\subsubsection{The HI detection rate}

It may also be a concern that the ALFALFA detects only gas rich galaxies at high redshift, and this will underestimate the group HI mass and introduce bias into our analysis.  We plot the HI detection fraction of the SDSS groups as a function of redshift, as shown in Fig.2. In Fig.2 the blue squares represent the HI detection rates of the spiral galaxies in the SDSS groups and each square represents the average over a velocity range of $\rm 500\;km\;s^{-1}$.  The black triangles represent the HI detection rates among all the member galaxies in the SDSS groups. It can be seen that when the group velocity is less than $\rm 8000\;km\;s^{-1}$, about 60\%-80\% spiral galaxies are detected by the ALFALFA while for groups with velocity greater than $\rm 8000\;km\;s^{-1}$ the HI detection rates drop significantly to 40\% and lower. The number of spirals and HI detections in each SDSS group are listed in Table A1.

If we include early type galaxies, the detection rate for the whole group is generally lower than 40\% and can be as low as 20\% for  velocity greater than $\rm 10000\;km\;s^{-1}$. This is not surprising, as the studies of \cite{di Serego Alighieri2007} and \cite{Grossi2009} found that the HI detection rate of early type galaxies is about 25\% in low density environments and 2.3\%  in high density environments such as the Virgo cluster, respectively.

\subsubsection{The detction limits and HI mass estimates in the SDSS groups }

For the groups containing  HI un-detected members,  we have estimated the possible missed HI mass based on the ALFALFA detection limits. We compute the minimum detectable HI mass at different distances using equation (5) of \cite{Giovanelli2005} assuming that a velocity width of $\rm 200\;km\;s^{-1}$ and an integration time of 40 seconds, but we use $4.5\sigma$  instead of $6\sigma$ in that equation. The  resulting detection limit of HI mass is  $4.1 \times 10^8 M_{\odot}$ at z=0.02 and $1.6 \times 10^9 M_{\odot}$ at z=0.04. Here $4.5\sigma$ is used because the ALFALFA catalog we used  is down to the limit of $4.5\sigma$. By summing up all the $4.5\sigma$ detection limit HI masses for the HI un-detected galaxies, and combined with the ALFALFA detected HI mass, we can estimate the upper limits of the HI masses for our sample SDSS groups.


Instead of using the above mentioned upper limits, we have also tried to obtain a better estimate of the HI masses for the  HI un-detected galaxies (including all morphological types) using the relationships between galaxy gas contents and optical colors. We first estimated the stellar masses from the r-band luminosity and g-r color using the formula derived by \cite{Bell2003}:
\begin{equation}
  \log(M_*/L_r)=-0.306+1.097(g-r)
\end{equation}
and then estimated the HI masses using the relationship between $M_{\rm HI}/M_{*}$ and g-r color which is fitted from Figure 2 of \cite{Zhang2009}, where $M_{*}$ is the galaxy stellar mass:
\begin{equation}
  \log(M_{\rm HI}/M_*)=1.09431-3.08207(g-r)
\end{equation}
The scatters of the two formulas are 0.1 and 0.35 dex, respectively. The r band absolute magnitude and g-r color of the member galaxies are listed in the SDSS Mr18 catalog. With these two formulas, we compute the HI mass of each HI un-detected member galaxy and the sum of these HI masses in one group is listed in Table A1 as $M_{\rm HI,color}$.
The total HI gas mass of a group is the sum of  $M_{\rm HI,color}$ and the ALFALFA detected HI mass $M_{\rm HI}$.


\begin{figure}
\centering
    \includegraphics[height=3.0in,width=3.2in,angle=0]{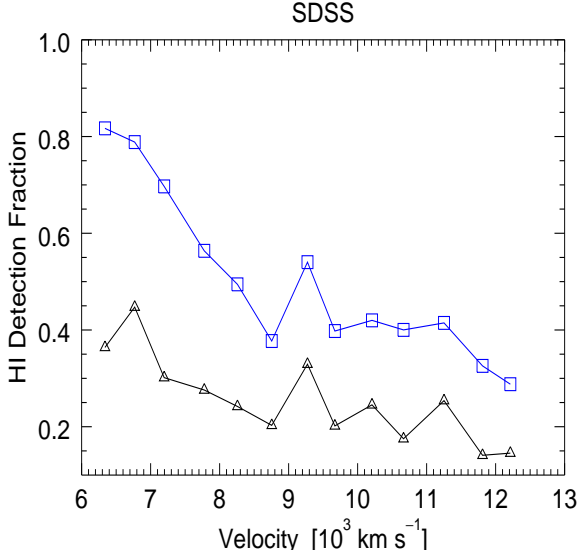}
    \caption{The HI detection fraction distribution of the SDSS groups as a function of group velocity.The blue squares represent the HI detections in all the spiral galaxies in the groups. The black triangles represent the HI detections in all the member galaxies in the groups.}
    \label{figure3}
\end{figure}

\subsection{Group virial Mass}

\begin{figure}
\centering
 \includegraphics[height=72mm,width=85mm]{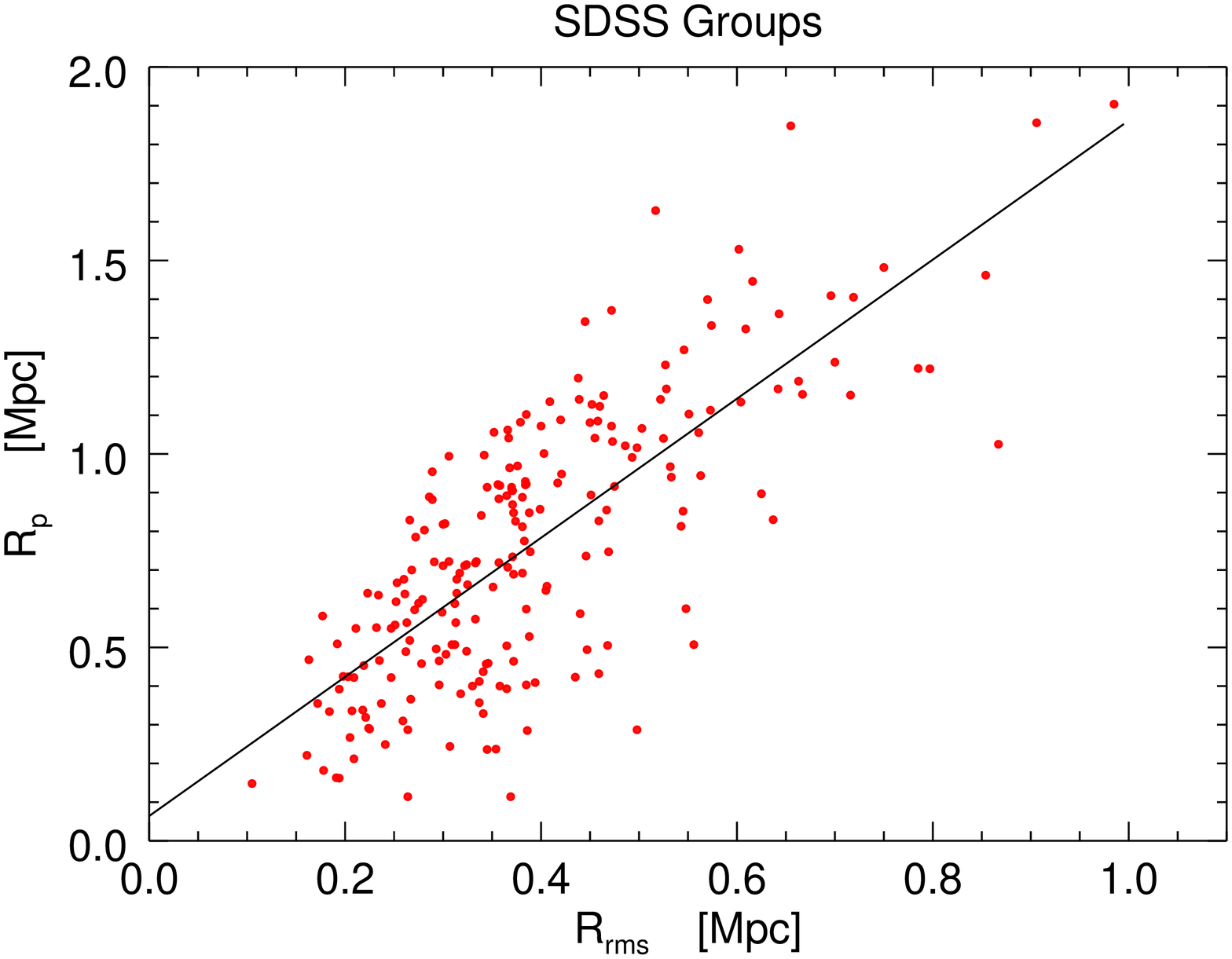}
 \caption{The correlation between $R_{\rm p}$ and $R_{\rm rms}$ in our SDSS sample. $R_{\rm p}$ is the projected virial radius derived in this work and $R_{\rm rms}$ is the rms projected radius listed in the SDSS group catalog. }
\end{figure}

\cite*{Dressler1988}  have proposed a method to test  if groups are virialized systems. Such method is called Dressler-Shectman (DS) test,  which compares the local velocity and velocity dispersion for each group member galaxy with the global group values to test the presence of substructures in clusters or groups. We have performed the DS test for our sample groups. For each galaxy i, we select its N nearest neighbors and compute their mean velocity $v_l$ and velocity dispersion $\sigma_l$. We compute the $\delta$ value for each of the SDSS groups as follows:
 \[ \delta^2=\frac{(N+1)}{\sigma_{group}^2} [(v_l-v_{group})^2+(\sigma_l-\sigma_{group})^2]\] where $v_{group}$ is the SDSS group mean velocity and $\sigma_{group}$ is the group velocity dispersion.
For the groups with number of member galaxies $N_{mem}$ greater than 20, we set N=10, otherwise $N=\sqrt{N_{mem}}$. The $\bigtriangleup$ value is $\bigtriangleup=\Sigma_i\delta$. A group is considered to be virialized (or do not have substructure) if  $\bigtriangleup/N_{mem} < 1 $.

Our results indicate that all of our sample groups do not have significant substructures, thus it is reasonable to assume that they are at least quasi-virialized systems.

The virial mass $M_{\rm v}$ is not listed in the SDSS Mr18 catalog.  Thus we calculate the group virial mass following equation (6) of \cite{Crook2008},  e.g.
\begin{equation}
  M_{\rm v}=3/2\pi \sigma^2_{\rm p}R_{\rm p}/G ,\qquad
  R_{\rm p}=N(N-1)/\sum_{i>j}R^{-1}_{ij}
\end{equation}
where $R_{ij}$ is the projected separation between two galaxies. We calculate $R_{\rm p}$ for every SDSS group and use it to compute $M_{\rm v}$. Fig.3 compares the values of $R_{\rm rms} $ and $R_{\rm P}$ for our sample groups and we found that these two quantities are closely correlated with each other. A straight line fitting yields a relation  $R_{\rm P}$$\sim$$1.8R_{\rm rms}$, with a scatter of $\sim$ 0.25, which corresponds to a relative error of 47\% in the group radius.

\subsection{SDSS group crossing time}

The group crossing time of the SDSS groups are calculated using the following formula similar to equation (3) of \cite{Tully1987}:
\begin{equation}
    t_{\rm c}=\frac{1.51^{1/2}R_{\rm rms}}{3^{1/2}\sigma_{\rm p}}
\end{equation}
where $R_{\rm rms}$ is the projected group radius defined in equation (8) of \cite{Berlind2006}, and $\sigma_{\rm p}$ is the velocity dispersion. The ratio of the crossing time to the approximate age of the Universe, $1/H_0$, is a convenient measure of the dynamical state of a group. Thus the parameter $t_{\rm c}H_{\rm 0}$ indicates the rough time that a galaxy traversed the group, and its reciprocal is the maximum number of times a galaxy could have traversed the group since its formation \citep{Hickson1992}. A smaller value of $t_{\rm c}H_{\rm 0}$ corresponds to a late evolutionary stage for a group. According to \cite{Berlind2006}, the velocity dispersion of the SDSS Mr18 groups is systematically underestimated by 20\%. We correct the velocity dispersion bias by applying an upward 20\% correction and use the corrected values to calculate the virial mass and crossing time.

\section{Results}

Numerical simulations have predicted that HI gas content in galaxies continually decreases in the course of group evolution, either by conversion to stars or by depletion due to various environmental effects \citep{Duffy2012}. In this section we study the relation between group HI mass fraction  $f_{\rm HI}=M_{\rm HI}/M_{\rm v}$ or group spiral fractions and group crossing time $t_{\rm c}$, where $M_{\rm v}$ is the group virial mass.

\subsection{Correlation between group spiral fraction and crossing time}
In section 2.2 we have obtained the spiral fraction and crossing time for each of the SDSS Mr18 groups. In Figure 4 we plot the group spiral fraction as a function of group crossing time. The green points represent the groups whose spiral galaxies are classified from Galaxy Zoo 1 with a quality cut of 0.5. It shows a weak  correlation with a correlation coefficient of 0.26 and the corresponding  statistical confidence level is 99.99\% (if a quality cut of 0.6 is used, the correlation coefficient is 0.25 and the confidence level is 99.99\% ). To reduce the influence due to uncertainty in the crossing time, we binned the data over a range of  $\log(t_{\rm c}H_0$)= 0.15.  The blue diamonds represent the groups whose spiral galaxies are classified from Galaxy Zoo 1 with a quality cut of 0.5, and the red diamonds represent those with a quality cut of 0.6. The red diamonds are systematically lower than the blue ones, but they share the same trend. Each diamond represents the averaged value and the error bars are the 1$\sigma$ standard deviation in each bin. The group spiral fractions show a clear trend of increasing with crossing time. For the cut value of 0.5, the correlation coefficient is 0.98 for the binned data and the corresponding confidence level is 99.99\% (for the cut value of 0.6 the correlation coefficient is 0.97 and the confidence level is 99.99\%).  This correlation is consistent with the result obtained from hyper compact groups (Fig.5, \citealt{Hickson1992}). Since spiral galaxies are usually rich in HI gas, we expect that the group HI gas fraction should increase with crossing time as well.

The above analysis is based on the mean value of the spiral fraction in each bin. We have also made similar analysis using the median value for each data bin and arrived at similar statistical results.  We have calculated the standard error of the median for each bin with a bootstrapping technique, and found that they are all less than 0.1 which is much smaller than the standard deviation in the data bin. Hence we consider that the errors on the mean or on the median values would have a limited effect on the statistical results in Fig.4.

\begin{figure}
\centering
    \includegraphics[height=3.0in,width=3.2in]{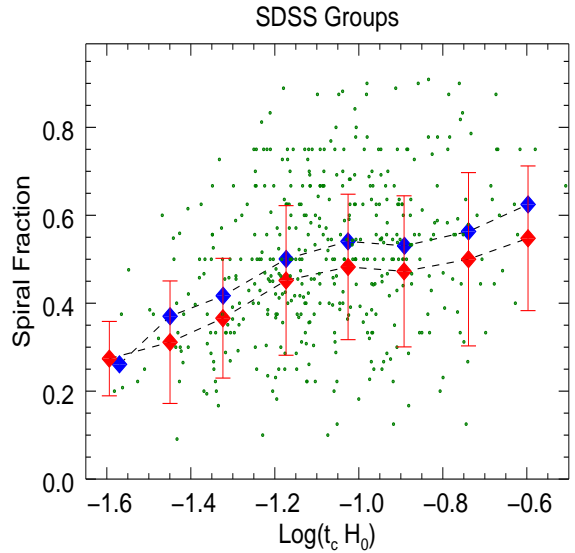}
    \caption{The SDSS group spiral fraction as a function of crossing time. The green points represent the groups whose spiral galaxies are classified from Galaxy Zoo 1 with a quality cut of 0.5. The blue diamonds are the binned value of the green points with a bin size of 0.15 classified with a quality cut of 0.5. The red diamonds are binned from those classified with a quality cut of 0.6.}
\end{figure}

\subsection{Correlation between group HI mass fraction and crossing time}

Before studying the correlation between $f_{\rm HI}$ and the group's crossing time $t_{\rm c}H_{\rm 0}$, we note that the denominator of $f_{\rm HI}$, $M_{\rm v}$, scales with $t_{\rm c}$ by definition where $t_{\rm c}$ is proportional to 1/$\sigma_{\rm p}$, and $1/M_{\rm v}$ is proportional to $(1/\sigma_{\rm p})^2$. To eliminate such an effect, we divide our sample into several subsamples according to the group virial mass so that $M_{\rm v}$ does not correlate with $t_{\rm c}H_{\rm 0}$ in each subsample. We exclude the groups with $M_{\rm v}$ less than $\rm 10^{13}\;M_{\odot}$ because the resulting subsample would become too small for a statistical study.

We split the SDSS sample into three subsamples, SDSS-1,SDSS-2 and SDSS-3, which correspond to the virial mass range of $\rm 10^{13}-10^{13.5}\;M_{\odot}$,$\rm 10^{13.5}-10^{14}\;M_{\odot}$ and $\rm 10^{14}-10^{14.5}\;M_{\odot}$. The upper panels of Fig.5 show the relation between $1/M_{\rm v}$ and  $t_{\rm c}H_{\rm 0}$.  The correlation coefficients between $1/M_{\rm v}$ and $t_{\rm c}H_{\rm 0}$ of the SDSS subsamples are listed in Table 1 as r1 which are all less than 0.3. The
statistical significance of r1 are shown as $\alpha 1$ in Table 1.  So, after narrowing down the virial mass range of each subsample there is almost no correlation between $1/M_{\rm v}$ and $t_{\rm c}H_{\rm 0}$  in each subsample. Thus any correlation between $f_{\rm HI}$ and  $t_{\rm c}H_{\rm 0}$, if it exists, would reflect the relation between the HI mass fraction and the group crossing time.

\begin{figure*}
\centering
 \includegraphics[height=2.5in,width=6.5in,angle=0]{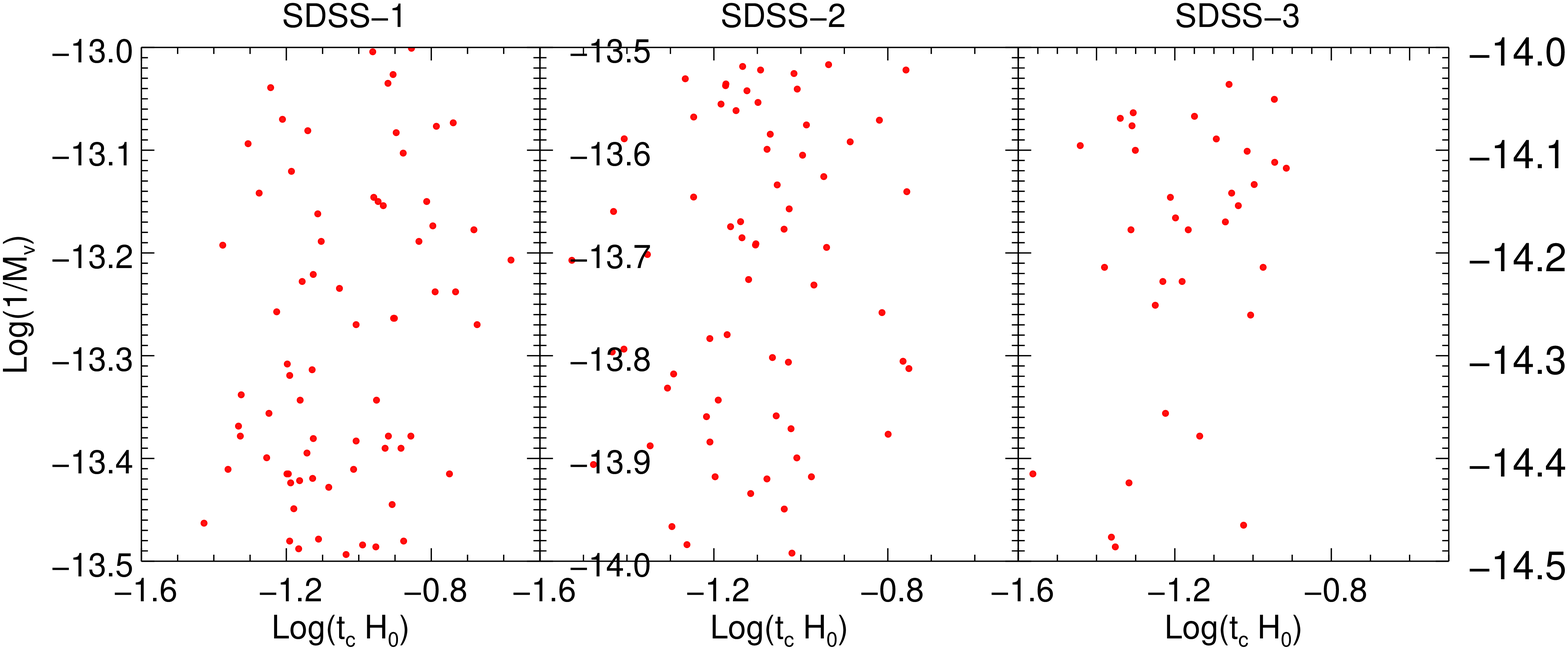}
 \includegraphics[height=2.5in,width=6.5in,angle=0]{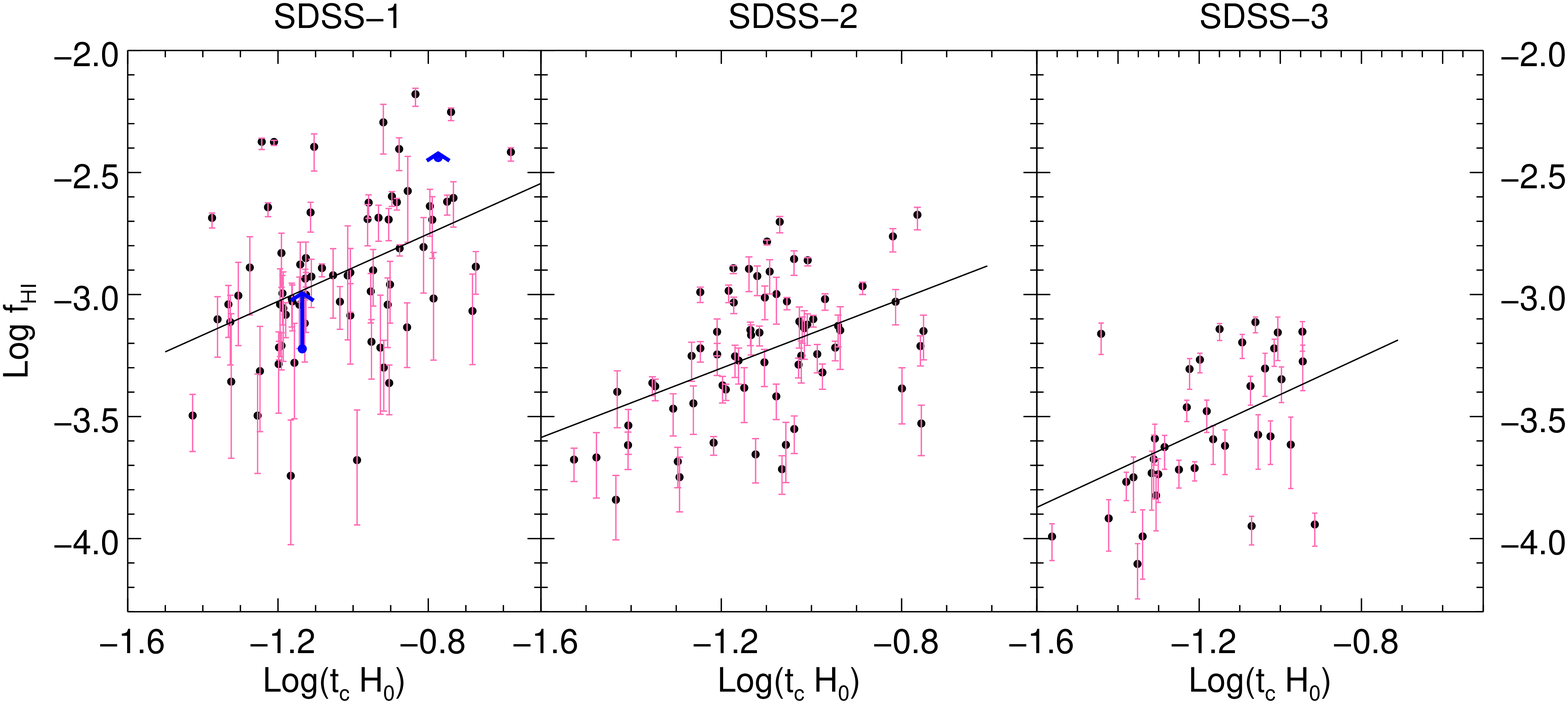}
 \includegraphics[height=2.5in,width=6.5in,angle=0]{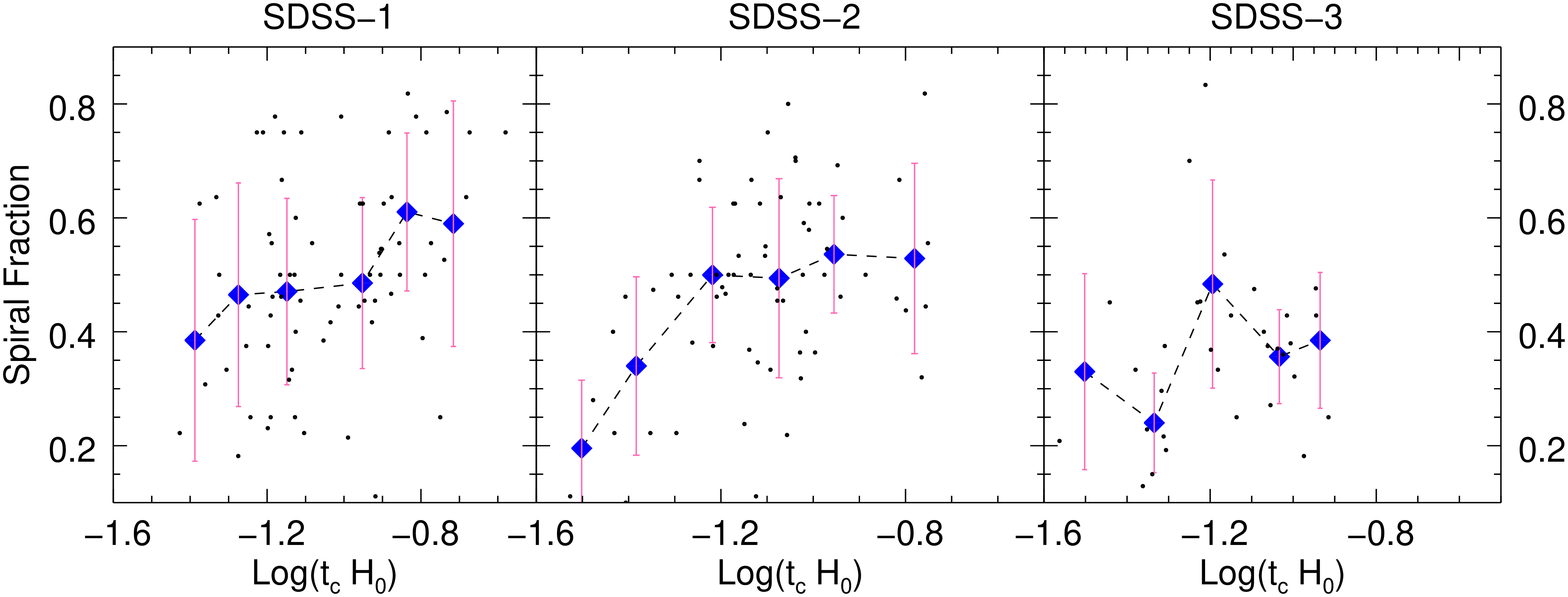}
 \caption{Correlation between the HI mass fraction and the crossing time (middle panels). The black line in each panel is the MCMC fit line. The upper panels show the correlation between the reciprocal of the virial mass and the crossing time for the SDSS groups. The middle panels show the correlation between spiral fractions and crossing time. The black points represent each group. The blue diamonds are the averages over $t_{\rm c}H_{\rm 0}$ of 0.15 (in log unit). The pink bars are the 1 $\sigma$ standard deviation. The two blue arrows in the middle panel represent the groups whose 4.5$\sigma$ upper limits are under the lower end of the error bar of the color estimated HI mass.}
\end{figure*}


\begin{table*}
\centering
\caption{The fitted parameters of a linear relation $y=ax+b$ between HI gas fraction and group crossing time.}
\renewcommand{\thetable}{\Alph{table2}}
 \begin{tabular}{lcccccccccccc}
   Subsample &$r$ & $a$  & $b$  & $r_{\rm 1\sigma}$ & $a_{\rm 1\sigma}$ & $b_{\rm 1\sigma}$  & $\alpha \le 0.03$ & r1 &$\alpha$1 &$<M_{\rm v}>$\\
              &  &    &    &      &    & & \% &  && $10^{13} \rm M_{\rm \odot}$\\
  \hline\noalign{\smallskip}
   SDSS-1    &0.41 &0.69  & -2.20 &0.031   &0.049   &0.052   &99.99    & 0.24 & 0.037  &2.02\\
   SDSS-2    &0.43 &0.71  & -2.45 &0.035   &0.053   &0.059   &99.99    & 0.15 & 0.236  &5.52 \\
   SDSS-3    &0.45 &0.77  & -2.64 &0.045   &0.092   &0.110   &93.44    & 0.25 & 0.159  &16.23\\
\end{tabular}
 \tablecomments{
Column 1: The subsample name;
Column 2: the correlation coefficient between $f_{\rm HI}$ and $t_{\rm c}H_{\rm 0}$;
Column 3: the slop of the fitted line;
Column 4: the intercept of the fitted line;
Column 5: the 1 $\sigma$ standard deviation of r;
Column 6: the 1 $\sigma$ standard deviation of a;
Column 7: the 1 $\sigma$ standard deviation of b;
Column 8: the percentage of fits that have a significance of the Spearman rank correlation $r$ less than 0.03;
Column 9: the correlation coefficient between $1/M_{\rm v}$ and $t_{\rm c}H_{\rm 0}$;
Column 10: the significance level of $r1$;
Column 11: the averaged virial mass.
 }
\end{table*}

In the middle panels of Fig. 5 we plot $f_{\rm HI}$  vs $t_{\rm c}H_{\rm 0}$  in logarithmic scale for each subsample. The group HI masses are the sums of $M_{\rm HI}$ and  $M_{\rm HI,color}$ which are all listed in Table A1.  The error bar for each group is derived from the scatter of Eq.(1) and (2) plus the observational error of the $\alpha.70$ HI detections.

We fit a straight line $y=ax+b$ to the points on the middle panels of Fig.5, where $x$ and $y$ represent crossing time $t_{\rm c}H_{\rm 0}$ and the HI gas fraction $f_{\rm HI}$ in logarithmic scale, respectively. We employ the Markov chain Monte Carlo (MCMC) method for the fitting.  For the HI mass of each group, we select a random value (assuming Gaussian distribution) within the HI mass error bars, with $M_{\rm HI,obs}$ + $M_{\rm HI,color}$ as the expected value. Note that we need to make sure that each of the total HI masses
used in the fitting is under the ALFALFA 4.5$\sigma$ upper limits discussed in section 2.3.4. We found that there are 147 groups whose 4.5$\sigma$ upper limits from the $\alpha.70$ are inside the error bars of the color estimated HI mass. In such case, during the MCMC process, we set the Gaussian probability distribution to zero beyond the 4.5$\sigma$ upper limit. This will help to avoid overestimating the group gas content with the color estimated HI mass. There are two groups whose 4.5$\sigma$ upper limits are under the lower end of the error bar of the color estimated HI mass. These two groups are shown as blue arrows in the middle panel of Fig.5. In the MCMC fitting for them we choose a random number based on uniform distribution in the range between the observed values and the 4.5$\sigma$ HI upper limits.

The errors in the virial mass and crossing time are mainly caused by uncertainties in group radius and velocity dispersion. As discussed in Sec 2.4, the relative error of the SDSS group radius is 47\%. It is difficult to estimate the error of the velocity dispersion of our SDSS groups, while according to \cite{Tully2015} the error of the group velocity dispersion is relatively small. For the groups with velocity dispersion between about 200-600 $\rm kms^{-1}$, the errors are less than 30\%. So we use an average value of 20\% for the  relative error of the velocity dispersion for the SDSS groups. Comparing to the group radius error, the error in the velocity dispersion is rather small and has very limited effect on our results. According to the error propagation formula, $t_{\rm c}H_{\rm 0}$ would have a relative error of $\sqrt{e_{\sigma}^2+e_{R}^2}$, and $M_{\rm v}$ would have a relative error of $\sqrt{4e_{\sigma}^2+e_{R}^2}$ where $e_{\sigma}$ and $e_{R}$ are the relative errors of group velocity dispersion and radius, respectively.

For each fit, we select a random value based on a Gaussian distribution for the $t_{\rm c}$ and $M_{\rm v}$ relative errors. Using the least $\chi^2$ fitting to the random values generated from the three categories of distribution ($M_{\rm HI}$,$t_{\rm c}H_{\rm 0}$ and $M_{\rm v}$), and repeating this process 100,000 times, we obtain three sets of distribution of the fit parameters: $a$, $b$ and $r$. The parameter $a$ is the slope, $b$ is the intercept of the fitted line and $r$ is the Pearson Correlation Coefficient for all the subsamples. The distributions of the parameters obtained from the 100,000 fits are Gaussian and we list the expectation value and 1$\sigma$ error of $r$, $a$ and $b$ in Table 1. All the values of $r$ are greater than 0.3 which means that these two variables are at least moderately correlated. Fig. 6 is an example of the distribution $r$ for subsample SDSS-2. Column 8 of Table 1 marked as $\alpha< 0.03$ shows the percentage of fits among the 100,000 repetitions that have a significance of the Spearman rank correlation less than 0.03 (corresponds to a confidence level of 97\%). We see that more than 90\% of the fits have $\alpha< 0.03$.

The analysis above is based on the assumption that the groups are all virialized so that the group virial mass can be calculated with Eq (3). If the group is not virialized, the group halo mass could be different from the virial mass. To estimate the uncertainties in $M_{\rm v}$  for our sample groups,  we have cross matched our group sample with the X-ray detected cluster catalog MCXC \citep{Piffaretti2011},  but only 27 SDSS Mr 18 groups are found.  Comparing our derived group virial masses with that derived from X-ray observations,  we found that the uncertainty in the virial mass could be by a factor of 2-3  in the worst cases.  Even taking such a large uncertainty into account,  our MCMC simulations show that  the correlation coefficients between $f_{HI}$ and $t_c$ remain larger than 0.3 in all subsamples, though they are slightly lower than  that listed in Table 1 as expected.

For comparison we also plot the spiral fraction distributions with $t_{\rm c}H_{\rm 0}$  for the three SDSS subsamples with the same virial mass ranges, as shown in the lower panels of Fig.5. The black points represent the morphological data obtained in Section 2.2.  We also binned the data over a range of $log t_{\rm c}H_{\rm 0}$ = 0.15. The blue diamonds represent the averaged spiral fraction of SDSS groups in the $t_{\rm c}H_{\rm 0}$ range. The pink bars are the 1 $\sigma$ standard deviation. From the middle and lower panels of Fig.5 we can see that the group spiral fraction is correlated with the crossing time following a trend similar to that of the HI gas fraction. The correlation parameters between spiral fraction and $t_{\rm c}H_{\rm 0}$ for all the SDSS subsamples (the lower panels of Fig.5) are listed in Table 2.
However, the fact that both $f_{\rm HI}$ and spiral fraction follow the same trend of decreasing with crossing time suggests that both gas fraction and morphology change as groups evolve, but the order in which this transformation takes place is unclear.

\begin{table*}
\centering
\caption{The correlation parameters between spiral fraction and $t_{\rm c}H_{\rm 0}$ for the SDSS subsamples.}
\renewcommand{\thetable}{\Alph{table2}}
 \begin{tabular}{c|c|c|c}
   Subsample &SDSS-1 & SDSS-2 & SDSS-3 \\
  \hline\noalign{\smallskip}
   un-binned    &0.4 \quad 99.7\%  & 0.3 \quad 98.9\%   &0.3 \quad  89.2\% \\
   binned       &0.9 \quad 99.5\%  & 0.9 \quad 98.1\%   &0.6 \quad  71.6\% \\
\end{tabular}
 \tablecomments{
  The upper row shows the correlation results for the un-binned data and the lower row shows the correlation results of the binned data in the lower panels of Fig.5. Each column lists the correlation coefficients and the confidence level for each subsample.
 }
\end{table*}

Table 1 also lists the average virial mass $<M_{\rm v}>$ of each subsample in units of $\rm 10^{13}\;M_{\odot}$ (column 11). From Fig.5 we can see that when the average virial mass increases, the group HI mass fraction decreases, and the MCMC fitted line is shifted toward lower $f_{\rm HI}$. Such effect is predicted in the numerical simulations(\citealt{Duffy2012}, \citealt{Dave2013}, \citealt{Cunnama2014}, \citealt{Rafieferantsoa2015}), and this is another reason why we needed to divide our sample into different subsamples.

The quantitative relation between $f_{\rm HI}$ and $t_{\rm c}$ from the MCMC fit is

\begin{equation}
 \log f_{\rm HI} = a\;\log(t_{\rm c}H_0)+b
\end{equation}

or

\begin{equation}
 f_{\rm HI} \propto t_{\rm c}^{a}\propto (R_{\rm rms}/\sigma_{\rm p})^{a}
\end{equation}
The slope $a$ is around 0.7 for each subsample.  If this correlation is due to the relation between $1/M_{\rm v}$ and $t_c$, the  slope should be about 0.5. Thus the increase of $f_{\rm HI}$ is most likely due to the increase of $M_{\rm HI}$ vs. crossing time.

\begin{figure}
\centering
 \includegraphics[height=3.0in,width=3.2in,angle=0]{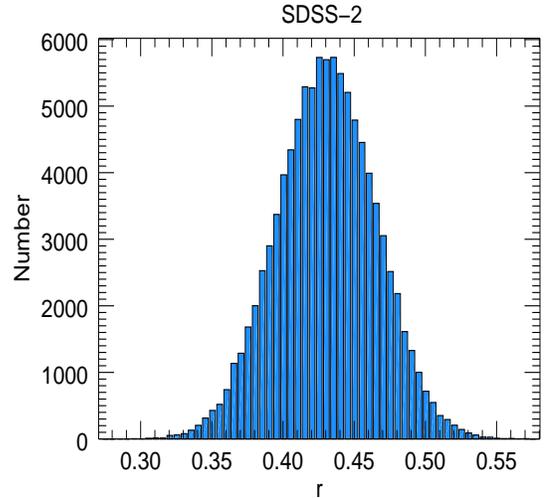}
 \caption{The Pearson Correlation Coefficient distribution after 100,000 fitting steps for the SDSS-2 subsample, bin size=0.005. }
\end{figure}

\subsection{Correlation between group HI mass fraction and group richness}

It has long been known that galaxy morphologies are related with environments following the morphology-density relation (T-$\Sigma$ relation)  (\citealt{Dressler1980}, \citealt{Cappellari2011}). In this section we explore the relation between the HI gas content and galaxy richness in groups. The upper left panel of Fig.7 shows the $f_{\rm HI}$ distribution as a function of the group member counts, where we use the group member counts N to represent the group richness, and  the data are binned together so that each bin contains a similar number of groups.  The blue diamond represents the averaged HI fraction $f_{\rm HI}$ in each data bin, for which the group HI masses are obtained from the  $\alpha.70$ catalog ($M_{\rm HI}$),  while the green diamonds represent those groups whose HI masses are calculated as $M_{\rm HI}+M_{\rm clolr,HI}$.   The upper right panel of Fig.7 shows the group spiral fraction as a function of group member counts.  Both plots show a clear trend of decreasing HI gas fraction and spiral fraction in rich groups with more galaxy members.
This result is  consistent with the T-$\Sigma$ relation, as rich groups have a higher number density environment.

Groups with more galaxy members also have a larger halo mass. We have divided our sample into 3 subsamples to remove part of the dependence on halo mass for the statistical studies.
The middle and lower panels of Fig.7 show the HI fraction $f_{\rm HI}$ and spiral fraction distribution as a function of the group member counts for the SDSS 1-3 subsamples.  We re-bin the range of N in each subsample so that each bin contains a similar number of groups. Comparing with the upper left panel in which $f_{\rm HI}$ decreases with N, there is no clear trend for the HI fraction $f_{\rm HI}$  as N increase among each subsample. The average HI fraction in different subsamples indeed dreases from SDSS-1 to SDSS-3.  This suggests that the anti-correlation between $f_{\rm HI}$ and N seen in the upper left pannel of Fig. 7 is mainly due to the increase of group halo mass, not due to the increase of galaxy members. Such result implies that  mechanisms related to halo mass such as starvation rather than merger are the major process in depleting HI gas in groups.

It is interesting to see that the group spiral fraction shows a dependence on the group member count N (lower panel of Fig.7). Unlike the HI fraction, the spiral faction in each subsample is anti-correlated with N. This suggests that merging is the key process in galaxy morphology transformation in groups.

\begin{figure*}
\centering
  \includegraphics[height=60mm,width=80mm,,angle=0]{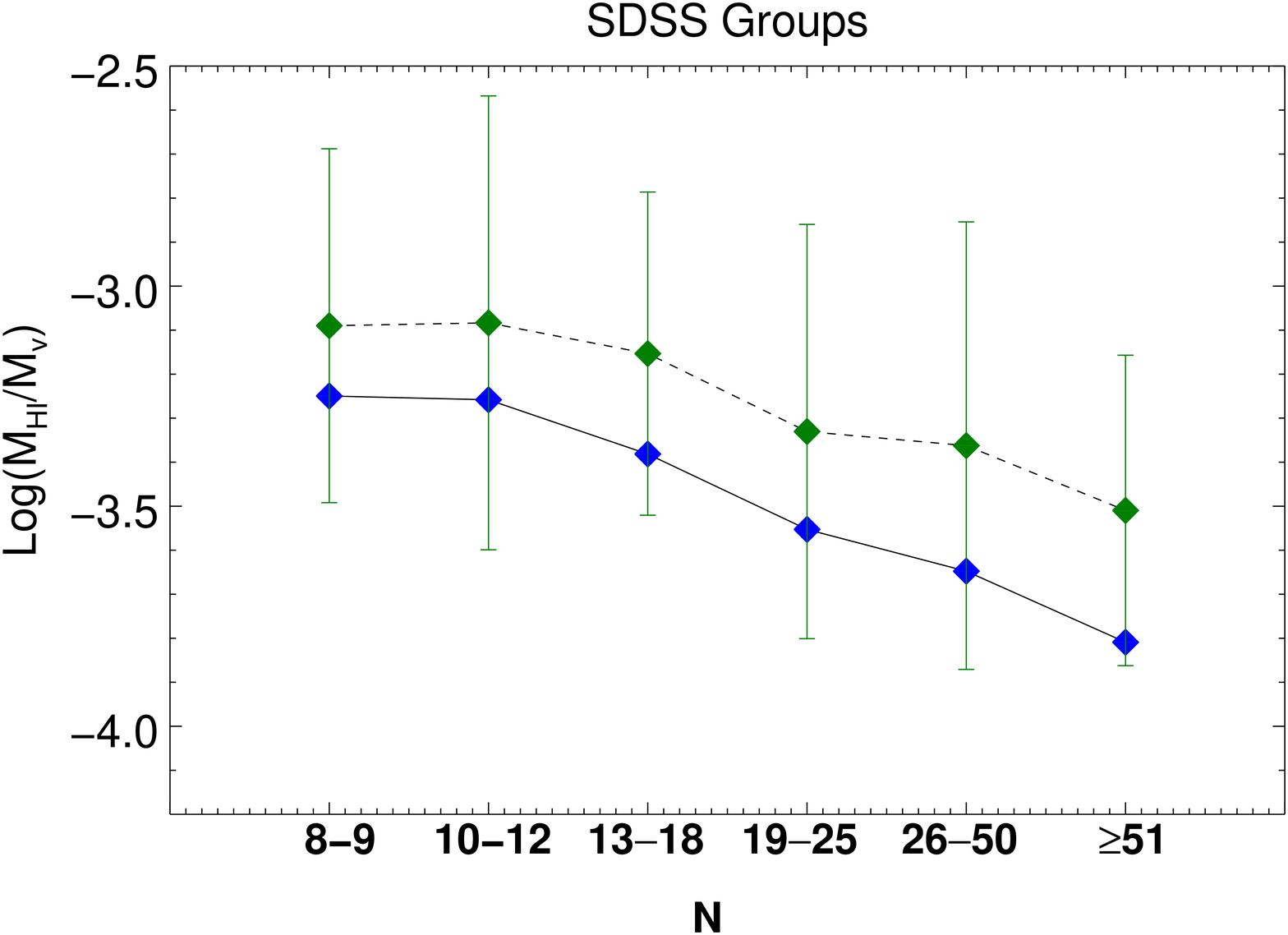}
  \includegraphics[height=60mm,width=80mm,,angle=0]{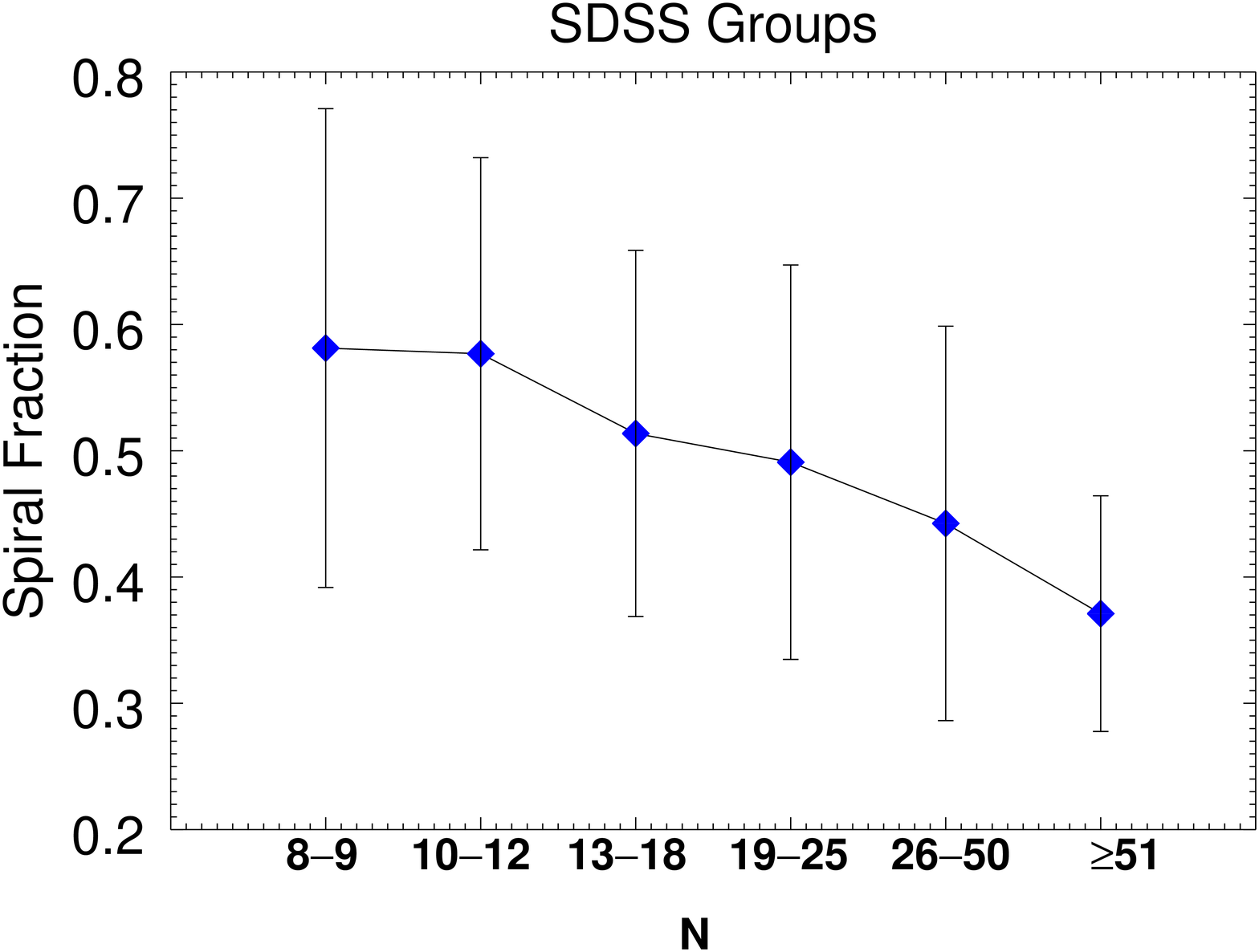}
  \includegraphics[height=2.5in,width=6.5in,angle=0]{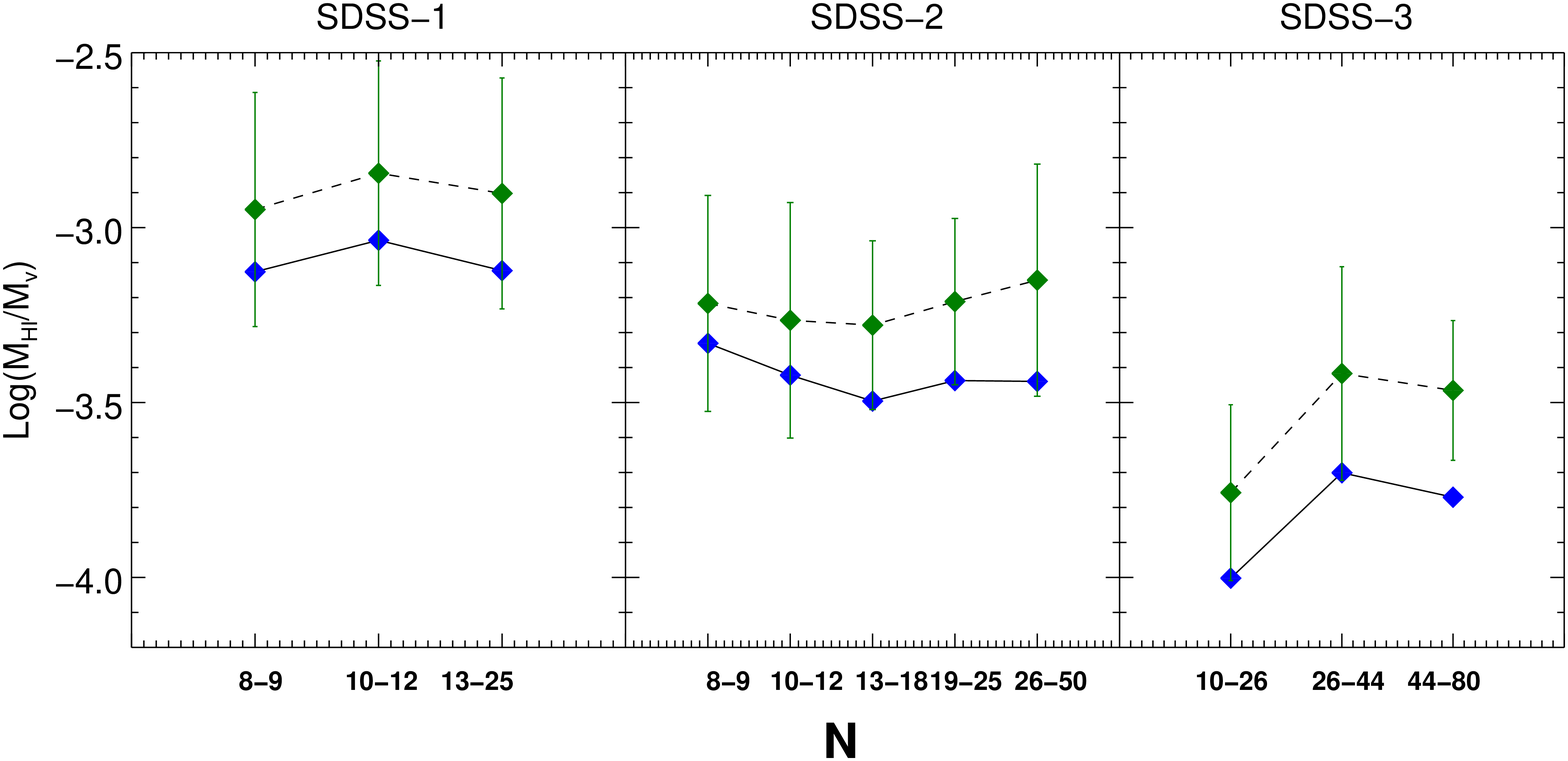}
  \includegraphics[height=2.5in,width=6.5in,angle=0]{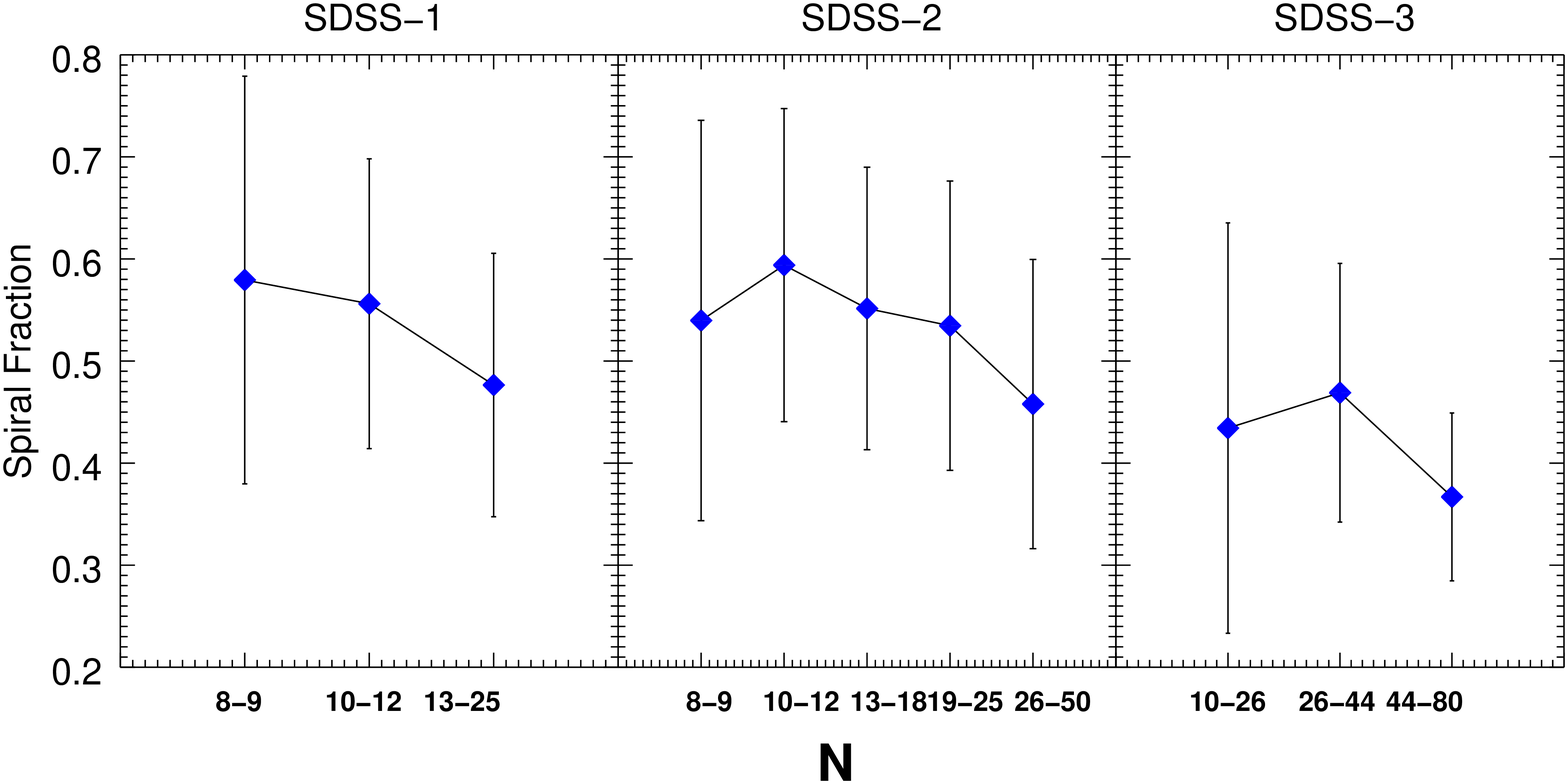}
 \caption{The correlations between group member counts with $f_{\rm HI}$ and spiral fractions. The upper left panel shows the $f_{\rm HI}$ distribution as a function of group member counts. The blue diamond represents the averaged HI fraction $f_{\rm HI}$ in each data bin, for which the group HI masses are obtained from the  $\alpha.70$ catalog ($M_{\rm HI}$),  while the green diamonds represent those groups whose HI masses are calculated as $M_{\rm HI}+M_{\rm clolr,HI}$.  The upper right panel shows the spiral fraction distribution as a function of group member counts. The error bars represent the 1 $\sigma$ standard deviation. The middle and lower panels show the same distribution as the upper panels but the sample is divided into three subsamples based on halo masses.}
\end{figure*}

\section{Discussion}

\subsection{The crossing time as an age indicator of galaxy groups}

Our purpose for studying the relation between  $f_{\rm HI}$ and $t_{\rm c}$  is to quantify the HI gas depletion process during the evolution of galaxy groups.  A key question is whether the crossing time $t_{\rm c}$  can be used as an indicator of group age.  This parameter was introduced by \cite{Tully1987} to
indicate the evolutionary state of galaxy groups. However, \cite{Diaferio1993} argued that
one should be cautious in using $t_{\rm c}$ as an indicator of group age, as his N-body simulation results showed that small groups (N $<$ 8) with small crossing times may be in the collapsed phase.
Thus we did not include groups with N $<$ 8 which could be dynamically young. We
have run the DS test for our SDSS groups and the test result shows that all our sample groups do not have significant substructures.   It is likely that the groups in our  sample are  at least quasi-virialized.
For a virialized system,   $M_{\rm v} \propto \sigma^2_{\rm p} R_{\rm p} $ and $M_{\rm v}=n R_{\rm p}^3$,   where n is the average density,we have $t_{\rm c} \propto R_{\rm p}/\sigma_{\rm p}$  $\propto 1/\sqrt {n}$. Thus, the value $t_{\rm c}$ is also related to the average density of the galaxy group.   As a group evolves to a later state, it gets  more and more compact and its average density increases.
From this perspective the crossing time $t_{\rm c}$ can indeed indicate the evolutionary state of a galaxy group.

\subsection{Relation with the "Butcher-Oemler" effect}
Evolutionary effects of galaxies are usually seen
at high redshifts, such as
the so called "Butcher-Oemler" effect (\citealt{Butcher1978},\citealt{Butcher1984}) according to which clusters
at z $>$ 0.4 have a substantial population of blue galaxies
implying more HI rich galaxies, while the nearby rich clusters
are very deficient in HI. However, very few clusters can be detected at HI at z $>$ 0.4, and thus we can not directly see the HI evolution at high z.
Nevertheless, using $t_c$, we can identify young galaxy groups (with large $t_c$) which could be the local analogs of the young groups at high z,
i.e.  large $t_c$ groups  at high z
could evolve into old groups with small $t_c$ at low z Universe.

More quantitatively,  \cite{Diaferio1993} estimated that 10 group crossing times would take about 5-6 Gyr in evolution.  Such change in terms of our definition of $t_cH_0$  is roughly from 0.04 to 0.3 (or -1.4 to -0.5 in logarithmic units) .  Hence the progenitor of a group with a small $t_c$ (e.g $0.04H_0^{-1}$) could have a larger value (about $0.3H_0^{-1}$) of $t_c$  at a "look back" time of 5 Gyr, or about z=0.5.  From Fig. 4, we can see that the spiral fraction almost doubles from crossing time of 0.04 to 0.3. Such morphology transformation due to $t_c$ is similar to the  B-O effect,  and  their time scales are well matched.  According to \cite{Moore1996} a dramatic morphological transformation in clusters which turns blue spiral galaxies to HI deficient elliptical galaxies occurred during a "look-back time" of about 4-5 billion years, consistent with about 10 group crossing times. We note that the  crossing time estimates of \cite{Diaferio1993}  are based on a simple numerical model. In the future, we plan to run detailed numerical simulations to model the change of $t_c$ of a group over the cosmic time and reveal its relation with the morphology transformation process.

Another effect that can reduce spiral fraction of a group is merging.  A small group can grow into a large group by accreting more galaxies over cosmic time, thus reduce the spiral fraction following the trends shown in Fig.7.
This process can also contribute to the B-O effect. Again detailed numerical simulations are required to reveal the contribution of different mechanisms in morphology transformation over cosmic time.

\subsection{The gas depletion time scale and environment effects}

Using the result obtained from Fig.5, we can make a more quantitative estimate of the HI depletion time in a group.  During the evolution of a galaxy group, after 10 crossing times,  e.g.  $t_cH_0$ increases from 0.04 to 0.3,   the  HI gas fraction  can decrease significantly. Thus the HI gas depletion time scale is approximately several Gyrs based on the time estimates in the previous section.

Ram pressure stripping in depleting HI gas is thought to be very efficient in galaxy clusters because the relevant time scale is very short ($\ge$ several $10^7$ yr, \citealt{Vollmer2012}). Thus this mechanism is unlikely the major effect for HI gas depletion in our sample groups.
On the other hand, the time scale of slowly acting processes such as starvation and strangulation is relatively long ($\ge$ 1 Gyr, \citealt*{Balogh2000}). Thus our results seem to fit with these long time HI depletion processes where gas experiences a steady decline.
Future theoretical modeling of HI gas evolution in galaxy groups taking into account all the environmental effects should be able to establish the accurate time scale for HI gas depletion and explain the observed relation between  $f_{\rm HI}$ and $t_{\rm c}$.


\section{Conclusions}

Our findings are summarized as the following:

1. By cross matching the latest released ALFALFA 70\% HI source catalog with the SDSS group catalog, we have identified 172 groups from the SDSS survey whose total HI mass can be derived by summing up the HI masses of all the HI sources within the group radius. As the HI gas is bound to the group and will eventually be the fuel  for star formation in the group, we treat the total HI gas content as a global quantity of the group and explore its relation with other group properties such as the virial mass. Based on the SDSS group sample we found a weak correlation between the group HI mass fraction and the group crossing time.

2. We used the Galaxy Zoo morphological data to derive the fraction of spiral galaxies in groups and  found a correlation between the group spiral fraction and the group crossing time for the SDSS Mr18 group sample. The galaxy morphology in our sample groups also follows the morphology-density relation.

3.  We suggest that the group crossing time is a good indicator of the group's age, and that younger groups with larger crossing times have more spiral galaxies as well as a higher fraction of HI gas than older groups.
The groups with large crossing times could be similar to the dynamically young systems at high redshift before they go through morphology transformation.

4. The detection rate of HI in the ALFALFA survey is still low for most groups with z$>$ 0.03. Several on-going HI surveys, such as CHILES with the VLA (COSMOS HI Large Extragalactic Survey, \citealt{Fernandez2016}), and LADUMA with the MeerKAT (Looking at the distant Universe with the MeerKAT Array, \citealt*{Blyth2016}, \citealt{Holwerda2012}), are designed to detect more high redshift HI signals, which are very important for understanding galaxy evolution.  Future deep HI surveys with large radio telescopes such as the FAST and SKA  will be able to improve the detection rate and improve the statistics on the total HI gas contents of groups and clusters.

5. Although the major focus of this paper is on the total HI gas in groups and did not include detailed analysis to distinguish the relative importance of the physical mechanisms for HI deficiency, the HI gas depletion time scale in our
sample groups are found to be several Gyrs,  similar to the time scale for morphology transformation.
This suggests that long time scale mechanisms such as starvation seem to play a more important role than short time scale processes in depleting HI gas in the SDSS galaxy groups.

\section{Acknowledgements}

We would like to thank the referee for his/her insightful comments and constructive suggestions. We also thank Dr. Martha Haynes for providing the ALFALFA data cubes of the SDSS groups in our sample. This work  was supported by the National Key $R\&D$ Program of China No. 2017YFA0402600 and by the NSFC grant No. U1531246. This work was also supported by the Open Project Program of the Key Laboratory of FAST, NAOC, Chinese Academy of Sciences. We acknowledge the work of the entire Galaxy Zoo team. Galaxy Zoo has been supported in part by a Jim Gray research grant from Microsoft, and by a grant from The Leverhulme Trust. We acknowledge the work of the entire ALFALFA collaboration team in observing, flagging, and extracting the catalog of galaxies used in this work. The Arecibo Observatory is operated by SRI International under a cooperative agreement with the National Science Foundation (AST-1100968), and in alliance with Ana G. Mendez-Universidad Metropolitana, and the Universities Space Research Association. Funding for SDSS-III has been provided by the Alfred P. Sloan Foundation, the Participating Institutions, the National Science Foundation and the US Department of Energy Office of Science.



\bibliographystyle{apj}

\begin{thebibliography}{}
\bibitem[Ai \& Zhu \& Fu(2017)]{Ai2017} Ai M.,Zhu M., Fu J., \ 2017, \ RAA, 17, 101
\bibitem[Balogh \& Navarro \& Morris(2000)]{Balogh2000} Balogh M.L.,Navarro, J.F., Morris, S.L., \ 2000, \apj, 540, 113
\bibitem[Balogh et al.(1997)]{Balogh1997} Balogh M.L., Morris S.L.,Yee H.K.C., et al. \ 1997, \apjl, 488, L75
\bibitem[Bekki \& Couch \& Shioya(2002)]{Bekki2002} Bekki,K., Couch, W. J., Shioya,Y., \ 2002, \apj, 577, 651
\bibitem[Bell et al. (2003)]{Bell2003} Bell, E.~F., McIntosh, D.~H., Katz, N., et al. \ 2003, \apjs, 149, 289
\bibitem[Berlind et al.(2006)]{Berlind2006} Berlind A.A., Frieman J., Weinberg D.H., et al. \ 2006, \apjs, 167, 1B
\bibitem[Bessell(1990)]{Bes90} Bessell, M.~S.\ 1990, \pasp, 102, 1181
\bibitem[Blanton \& Moustakas(2009)]{Blanton2009}  Blanton M.R.,  Moustakas J., \ 2009 \araa, 47,159B
\bibitem[Blyth(2016)]{Blyth2016} Blyth, S., Baker, A.~J., Holwerda, B., et al. \ 2016 \ mks, 4B
\bibitem[Boselli \& Gavazzi(2006)]{Boselli2006}  Boselli A.,  Gavazzi G., \ 2006, \pasp ,118,517B
\bibitem[Bretherton et al.(2013)]{Bretherton2013} Bretherton C.~F., Moss C., James P.A., et al. \ 2013,\aa,553A,67B
\bibitem[Brown et al.(2015)]{Brown2015} Brown, T., Catinella, B., Cortese, L., et al. \ 2015, \mnras, 452, 2479
\bibitem[Brown et al.(2016)]{Brown2016} Brown,G.M., Johnston. K.G., Hoare, M.G., et al. \ 2016, \mnras, 463, 2839
\bibitem[Butcher \& Oemler(1978)]{Butcher1978} Butcher H., Oemler Jr. A.,\ 1978, \apj, 219, 18
\bibitem[Butcher \& Oemler(1984)]{Butcher1984} Butcher H., Oemler Jr. A.,\ 1984, \apj, 285, 426
\bibitem[Cappellari et al.(2011)]{Cappellari2011} Capepellari M., Emsellem E., Krajnovic D., et al. \ 2011, \mnras, 416,1680C
\bibitem[Catinella et al.(2013)]{Catinella2013} Catinella B., Schiminovich, D., Cortese, L., et al. \ 2013, \mnras, 436,34C
\bibitem[Chung et al.(2009)]{Chung2009} Chung A., van Gorkom J.H., Kenney J. D.P., et al. \ 2009, \aj, 138,1741C
\bibitem[Cortese et al.(2011)]{Cortese2011} Cortese L.,  Catinella B., Boissier S., et al. \ 2011, \mnras,415, 1797
\bibitem[Crook et al.(2008)]{Crook2008} Crook A.~C.,  Huchra J.~P.,  Martimbeau N., et al. \ 2008 \apj, 685, 1320
\bibitem[Cunnama et al.(2014)]{Cunnama2014} Cunnama D.,  Andrianomena S.,  Cress., et al. \ 2014 \mnras, 438, 2530
\bibitem[Dave et al.(2013)]{Dave2013} Dave R., Katz, N., Oppenheimer, B.~D., et al. \ 2013 \mnras, 434, 2645
\bibitem[Davies \& Lewis(1973)]{Davies1973} Davies R.~D.,  Lewis B.M.,\ 1973, \mnras, 165, 231
\bibitem[di Serego Alighieri et al.(2007)]{di Serego Alighieri2007} di Serego Alighieri,S.,  Gavazzi, G. Giovanardi, C., et al. \ 2007 \aa, 474, 851
\bibitem[Diaferio et al.(1993)]{Diaferio1993} Diaferio,A., Ramela,M., Geller,M., et al. \ 1993 \aj, 105, 2035
\bibitem[Dressler(1980)]{Dressler1980} {Dressler} A., \ 1980, \apjs, 42, 565
\bibitem[Dressler \& Shectman (1988)]{Dressler1988} {Dressler} A., Shectman, S.~A.,\ 1988, \aj, 95, 985
\bibitem[Duarte \&  Mamon(2014)]{Duarte2014} Duarte M., Mamon G. A., \ 2014, \mnras, 440, 1763
\bibitem[Duffy et~al.(2012)]{Duffy2012} Duffy A.~R.,  Meyer M.~J.,  Staveley-Smith L., et al. \ 2012, \mnras, 426, 338
\bibitem[Fernandez et al.(2016)]{Fernandez2016} Fernandez, X., Gim, H.~B., van Gorkom, J.~H., et al. \ 2016, \ apj, 824, 1
\bibitem[Giovanelli \& Haynes(1985)]{Giovanelli1985} {Giovanelli} R.,  {Haynes} M.~P., \ 1985, \apj, 292, 404
\bibitem[Giovanelli et~al.(2005)]{Giovanelli2005} Giovanelli R., Haynes, M.~P., Kent, B.R., et al. \ 2005, \aj, 130, 2598
\bibitem[Grossi et al.(2009)]{Grossi2009} Grossi,M., di Serego Alighieri,S., Giovanardi, C., et al. \ 2009, \aa, 498, 407
\bibitem[Gunn \& Gott(1972)]{Gunn1972} Gunn J.~E.,  Gott J.~Richard I.,\ 1972, \apj, 176, 1
\bibitem[Haynes \& Giovanelli (1984)]{Haynes1984} Haynes M.~P., Giovanelli R.,\ 1984, \aj, 89, 758
\bibitem[Haynes et~al.(2011)]{Haynes2011} Haynes M.~P., Giovanelli, R., Martin, A.~M., et al. \ 2011, \aj, 142, 170
\bibitem[Hess \& Wilcots(2013)]{Hess2013} Hess K.~M.,  Wilcots E.~M.,\ 2013, \aj, 142, 170
\bibitem[Hibbard \& van Gorkom(1996)]{Hibbard1996} Hibbard,J.E.,  van Gorkom, J.H., \ 1996, \aj,111,655H
\bibitem[Hickson (1982)]{Hickson1982} Hickson P.,\ 1982, \apj,255,382
\bibitem[Hickson et~al.(1992)]{Hickson1992} Hickson P.,  Oliveira D.,  Claudia M., et al.\ 1992, \apj,399,353H
\bibitem[Holwerda \& Blyth \& Backer(2012)]{Holwerda2012} Holwerda, B.~W., Blyth, S.-L., Baker, A.~J., \ 2012, \ iaus, 283,496
\bibitem[Huang \& Kauffmann(2014)]{Huang2014} Huang, M.-L., Kauffmann, G., \ 2014, \mnras, 443, 1329
\bibitem[Kauffmann et~al.(2004)]{Kauffmann2004} Kauffmann G., White S.~D.~M., Heckman T.~M., et al. \ 2004, \mnras, 353, 713
\bibitem[Kawata \& Mulchaey (2008)]{Kawata2008} Kawata, D.,  Mulchaey,J.S., \ 2008 \apj, 672, 103
\bibitem[Kenney et~al.(2004)]{Kenney2004} Kenney J.~D.~P., van Gorkom J.~H., Vollmer B., et al. \ 2004,\aj,127, 3361
\bibitem[Kilborn et al.(2009)]{Kilborn2009} Kilborn,V. A., Forbes,D. A., Barnes,D.G., et al. \ 2009, \mnras, 400, 1962
\bibitem[Larson \& Tinsley \& Caldwell(1980)]{Larson1980} Larson R.B., Tinsley, B. M. Caldwell, C.N. \ 1980, \apj, 237, 692
\bibitem[Lewis et~al.(2002)]{Lewis2002} Lewis I., Balogh, M., De Propris, R., et al. \ 2002 \mnras,334, 673
\bibitem[Lintott et~al.(2011)]{Lintott2011} Lintott C., Schawinski, K., Bamford, S., et al. \ 2011, \mnras,410, 166
\bibitem[Mihos(2004)]{Mihos2004} Mihos J.~C., in Clusters of Galaxies: Probes of Cosmological Structure and Galaxy Evolution, ed. J. S. Mulchaey, A. Dressler, A. Oemler
(Cambridge: Cambridge Univ. Press), \ 2004, 277M
\bibitem[Mok et~al.(2016)]{Mok2016} Mok A., Wilson, C.~D., Golding, J., et al. \ 2016,\mnras, 456, 4384
\bibitem[Moore et~al.(1996)]{Moore1996} Moore B., Katz N., Lake G., et al. \ 1996, \ nature, 379, 613
\bibitem[Nan et al.(2011)]{Nan2011} Nan, R., Li,D., Jin, C. et al. \ 2011, \ ijmpd, 20, 989
\bibitem[Pasetto et~al.(2014)]{Pasetto2014} Pasetto S., Cropper M.,  Fujita Y.,et al. \  2014, \aa,573, A48
\bibitem[Piffaretti et al. (2011)]{Piffaretti2011} Piffaretti, R.,  Arnaud, M., Pratt, G.~W., et al. \ 2011, \aa, 534, 109
\bibitem[Rafieferantsoa et al.(2015)]{Rafieferantsoa2015} Rafieferantsoa M., Dav{\'e}, R., Angl{\'e}s-Alc{\'a}zar D., et al. \ 2015 \mnras, 453, 3980
\bibitem[Rasmussen et al.(2012)]{Rasmussen2012} Rasmussen,J., Mulchaey, J.S., Bai,L. et al. \ 2012, \apj, 757, 122
\bibitem[Saintonge et al.(2011)]{Saintonge2011} Saintonge, A., Kauffmann, G., Kramer,C. et al. \ 2011, \mnras, 415, 32
\bibitem[Saintonge et al.(2012)]{Saintonge2012} Saintonge, A., Tacconi, L.~J., Fabello, S. et al. \ 2012, \apj, 758, 73
\bibitem[Serra et al.(2012)]{Serra2012} Serra, P., Oosterloo,T. Morganti,R. et al. \ 2012, \mnras, 422, 1835
\bibitem[Skrutskie(2006)]{Skrutskie2006} Skrutskie M.~F.,\ 2006, \aj, 131, 1163
\bibitem[Solanes(2001)]{Solanes2001}Solanes, J.~M. Manrique, A. and Garc{\'{\i}}a-G{\'o}mez, C., et al. \ 2001, \apj, 548,97
\bibitem[Solanes et~al.(2004)]{Solanes2004} Solanes J.~M.,  Sanchis T.,  Salvador-Sol\'{e} E., et al. \ 2004, ogci,401
\bibitem[Stark et al.(2016)]{Stark2016} Stark, D. V., Kannappan,S.J. Eckert, D. D., et al. \ 2016, \apj, 832, 126
\bibitem[Tacconi et~al.(2013)]{Tacconi2013} Tacconi, L. J,,  Neri R.,  Combes, F. et al. \ 2013, \apj, 768, 74
\bibitem[Taylor et~al.(2012)]{Taylor2012} Taylor R.,  Davies J.~I.,  Auld R., et al. \ 2012, \mnras,423, 787
\bibitem[Tully(1987)]{Tully1987} Tully R.~B., \ 1987, \apj, 321, 280
\bibitem[Tully(2013)]{Tully2013} Tully R.~B., \ 2013, \aj, 146, 86
\bibitem[Tully(2015)]{Tully2015} Tully R.~B., \ 2015, \aj, 149, 54
\bibitem[Verdes-Montenegro et~al.(2001)]{Verdes-Montenegro2001} Verdes-Montenegro L.,  Yun M.~S., Williams B.~A., et al. \  2001, \aa, 377, 812
\bibitem[Vollmer et~al.(2012)]{Vollmer2012}Vollmer B., Wong O. I., Braine J., Chung A., Kenney J. D. P., 2012, \ aa,543, A33
\bibitem[Wen \& Han(2015)]{Wen2015} Wen, Z.L., Han, J.~L., \ 2015, \apj, 807, 178
\bibitem[York et~al.(2000)]{York2000} York D.~G.,Adelman, J., Anderson, Jr., et al. \ 2000, \aj, 120, 1579
\bibitem[Zhang et~al.(2009)]{Zhang2009} Zhang,W., Li,C., Kauffmann,G., et al. \ 2009, \mnras, 397, 1243
\end{thebibliography}

\newpage
\appendix
\section{A. Table of group properties}
\clearpage
\label{app:co2}
\setcounter{table}{0}
\renewcommand{\thetable}{A\arabic{table}}
 \begin{table*}
\label{table2}
\centering	
\caption{The resulting values for the SDSS sample. (continued on next page)}
\begin{tabular}{lllllllllllll}
\hline
groupID &ra &dec &z & Nm &$N_{\rm spiral}$ &$N_{\rm HI}$ &$M_{\rm HI}$ &$M_{\rm v}$ &$t_{\rm c}H_{\rm 0}$ &$M_{\rm HI,color}$ & $M_{\rm HI,error}$ \\
        &   &    &  &    &                 &             &$\rm 10^{10} M_{\odot}$ &$\rm 10^{13} M_{\odot}$ &              & $\rm 10^{9} M_{\odot}$ &$\rm 10^{8} M_{\odot}$     \\
\hline
6	&	38.172886	&	0.668691	&	0.02193	&	8	&	3	&	5	&	1.58 	&	7.23 	 &	0.061	&	2.07 	&	8.88 	\\
209	&	237.375259	&	0.277424	&	0.03225	&	8	&	2	&	1	&	1.02 	&	2.09 	 &	0.064	&	2.67 	&	7.46 	\\
265	&	164.568848	&	1.558856	&	0.03935	&	79	&	35	&	8	&	7.07 	&	18.20 	 &	0.099	&	56.60 	&	59.40 	\\
284	&	212.780563	&	1.367162	&	0.02524	&	11	&	6	&	2	&	1.20 	&	1.41 	 &	0.113	&	5.74 	&	6.80 	\\
464	&	162.506287	&	0.360061	&	0.03917	&	21	&	9	&	1	&	1.62 	&	9.64 	 &	0.055	&	18.30 	&	12.90 	\\
591	&	181.098618	&	1.689661	&	0.02066	&	14	&	5	&	1	&	0.20 	&	3.05 	 &	0.102	&	4.35 	&	1.95 	\\
700	&	29.357054	&	14.545851	&	0.0263	&	8	&	3	&	4	&	1.11 	&	2.03 	 &	0.063	&	1.24 	&	9.41 	\\
712	&	18.38389	&	15.724561	&	0.038	&	12	&	7	&	4	&	2.26 	&	1.73 	 &	0.162	&	12.40 	&	15.40 	\\
1910	&	228.058075	&	1.983612	&	0.03875	&	44	&	10	&	1	&	0.38 	&	 16.40 	&	0.106	&	35.90 	&	3.97 	\\
1932	&	218.560852	&	3.535023	&	0.02884	&	42	&	23	&	7	&	5.52 	&	 11.20 	&	0.113	&	23.90 	&	22.40 	\\
2002	&	196.161041	&	3.769489	&	0.0403	&	11	&	5	&	3	&	2.63 	&	 1.45 	&	0.077	&	5.20 	&	25.70 	\\
2005	&	220.170898	&	3.547091	&	0.027	&	59	&	17	&	3	&	1.29 	&	 13.90 	&	0.088	&	24.00 	&	7.50 	\\
2128	&	169.497421	&	2.609058	&	0.02958	&	8	&	5	&	2	&	0.93 	&	 2.20 	&	0.112	&	4.81 	&	6.21 	\\
2255	&	219.220154	&	3.36729	&	0.02659	&	9	&	2	&	1	&	0.41 	&	3.48 	 &	0.075	&	3.64 	&	3.00 	\\
2263	&	223.368011	&	3.20061	&	0.028	&	16	&	10	&	3	&	2.28 	&	6.02 	 &	0.068	&	10.80 	&	11.90 	\\
2268	&	228.990738	&	2.991235	&	0.03763	&	14	&	8	&	2	&	1.10 	&	 2.39 	&	0.047	&	7.43 	&	10.80 	\\
2293	&	215.777206	&	4.496596	&	0.02656	&	11	&	9	&	4	&	1.52 	&	 3.33 	&	0.174	&	5.25 	&	10.70 	\\
2326	&	223.403412	&	4.581405	&	0.0283	&	9	&	2	&	1	&	0.71 	&	 2.39 	&	0.121	&	4.91 	&	3.92 	\\
2332	&	227.996246	&	4.518596	&	0.03617	&	48	&	15	&	5	&	3.14 	&	 23.90 	&	0.073	&	25.90 	&	28.60 	\\
2336	&	229.562668	&	4.418047	&	0.03675	&	49	&	22	&	4	&	3.46 	&	 12.90 	&	0.114	&	34.20 	&	23.90 	\\
2752	&	348.358185	&	14.134232	&	0.03974	&	11	&	6	&	4	&	3.95 	&	 1.08 	&	0.12	&	15.50 	&	23.40 	\\
2760	&	351.260101	&	14.626508	&	0.04047	&	75	&	28	&	5	&	4.14 	&	 29.20 	&	0.095	&	35.10 	&	30.50 	\\
2800	&	333.635864	&	13.779809	&	0.02602	&	45	&	16	&	7	&	3.77 	&	 16.90 	&	0.066	&	18.50 	&	18.10 	\\
3480	&	167.594254	&	4.669164	&	0.02983	&	28	&	16	&	9	&	7.41 	&	 11.70 	&	0.071	&	10.10 	&	32.40 	\\
3510	&	135.683716	&	3.466104	&	0.02663	&	8	&	6	&	1	&	0.39 	&	 1.69 	&	0.07	&	4.98 	&	3.31 	\\
3562	&	159.694855	&	5.669923	&	0.02812	&	12	&	6	&	6	&	3.99 	&	 3.91 	&	0.13	&	2.36 	&	17.30 	\\
3620	&	168.751434	&	4.132484	&	0.03957	&	19	&	7	&	2	&	1.32 	&	 2.48 	&	0.072	&	9.34 	&	11.90 	\\
3943	&	206.886124	&	3.611311	&	0.02342	&	10	&	7	&	4	&	1.84 	&	 3.70 	&	0.057	&	3.84 	&	7.76 	\\
3962	&	203.434601	&	4.726468	&	0.02214	&	8	&	5	&	6	&	4.61 	&	 3.47 	&	0.098	&	1.85 	&	13.00 	\\
4005	&	205.21727	&	4.808541	&	0.02316	&	19	&	11	&	14	&	6.17 	&	 1.18 	&	0.182	&	4.51 	&	25.90 	\\
4068	&	188.851364	&	5.843061	&	0.0408	&	8	&	5	&	2	&	1.22 	&	 3.39 	&	0.054	&	6.81 	&	11.20 	\\
5375	&	232.004517	&	4.009582	&	0.03807	&	9	&	4	&	2	&	1.58 	&	 1.01 	&	0.109	&	4.77 	&	11.80 	\\
5424	&	226.504929	&	5.559979	&	0.03681	&	24	&	14	&	5	&	5.24 	&	 3.72 	&	0.152	&	12.00 	&	28.60 	\\
5447	&	233.069153	&	4.768817	&	0.03877	&	37	&	13	&	1	&	0.69 	&	 15.00 	&	0.049	&	24.90 	&	5.76 	\\
5502	&	214.607208	&	7.474541	&	0.02469	&	20	&	10	&	4	&	1.29 	&	 6.78 	&	0.049	&	10.20 	&	9.19 	\\
6059	&	3.154815	&	-0.021308	&	0.03967	&	9	&	2	&	3	&	4.93 	&	 1.54 	&	0.079	&	12.90 	&	16.50 	\\
7335	&	125.903854	&	4.291001	&	0.02849	&	10	&	4	&	1	&	0.21 	&	 6.26 	&	0.037	&	6.88 	&	2.74 	\\
7383	&	162.861298	&	8.505543	&	0.02163	&	15	&	7	&	7	&	2.58 	&	 6.97 	&	0.065	&	2.64 	&	13.20 	\\
7621	&	177.848831	&	8.761795	&	0.03539	&	9	&	4	&	1	&	0.42 	&	 2.27 	&	0.056	&	6.85 	&	4.29 	\\
7625	&	164.530182	&	8.48995	&	0.03502	&	9	&	4	&	2	&	1.32 	&	2.20 	 &	0.073	&	23.20 	&	9.50 	\\
7633	&	181.046585	&	9.184257	&	0.03479	&	9	&	5	&	1	&	1.10 	&	 2.39 	&	0.139	&	6.53 	&	5.43 	\\
7638	&	186.547516	&	8.892553	&	0.02448	&	15	&	9	&	5	&	1.73 	&	 4.73 	&	0.069	&	8.05 	&	15.00 	\\
7699	&	164.520813	&	9.42227	&	0.03413	&	11	&	6	&	3	&	1.30 	&	1.83 	 &	0.125	&	7.16 	&	10.80 	\\
7700	&	165.15062	&	9.401519	&	0.03589	&	18	&	7	&	3	&	2.52 	&	 1.49 	&	0.16	&	9.12 	&	13.90 	\\
7768	&	164.004959	&	9.750942	&	0.03237	&	8	&	6	&	5	&	3.81 	&	 1.81 	&	0.059	&	3.10 	&	19.90 	\\
7787	&	185.83316	&	10.63978	&	0.02527	&	13	&	6	&	1	&	0.40 	&	 6.57 	&	0.051	&	7.75 	&	2.05 	\\
7814	&	165.249725	&	10.463159	&	0.03632	&	22	&	8	&	3	&	2.01 	&	 4.54 	&	0.094	&	15.10 	&	13.90 	\\
9146	&	156.77887	&	11.003482	&	0.03216	&	28	&	17	&	3	&	2.60 	&	 6.07 	&	0.062	&	16.70 	&	13.30 	\\

\hline											
\end{tabular}
 \tablecomments{
Column 1: Group ID;
Column 2: R.A. (J2000) of group;
Column 3: Decl. (J2000) of group;
Column 4: redshift of group;
Column 5: number of member galaxies;
Column 6: number of spiral galaxies;
Column 7: number of galaxies detected by HI;
Column 8: group HI mass detected by the ALFALFA;
Column 9: group virial mass;
Column 10: crossing time;
Column 11:group HI mass of the HI-undetected member galaxies estimated based on optical color;
Column 12:the error of the detected HI mass of the SDSS groups.
 }
\end{table*}											
											
\addtocounter{table}{-1}											
\begin{table*}											
\centering											
\caption{The resulting values for the SDSS sample. (continued from previous page)}											
\begin{tabular}{lllllllllllll}											
\hline											
groupID &ra &dec &z & Nm &$N_{\rm spiral}$ &$N_{\rm HI}$ &$M_{\rm HI}$ &$M_{\rm v}$ &$t_{\rm c}H_{\rm 0}$ &$M_{\rm HI,color}$ & $M_{\rm HI,error}$ \\
        &   &    &  &    &                 &             &$\rm 10^{10} M_{\odot}$ &$\rm 10^{13} M_{\odot}$ &              &      $\rm 10^{9} M_{\odot}$       &$\rm 10^{8} M_{\odot}$  \\
\hline										
9168	&	148.512573	&	10.588834	&	0.04044	&	15	&	8	&	2	&	1.20 	&	 3.35 	&	0.096	&	12.40 	&	11.80 	\\
9428	&	167.008255	&	13.135915	&	0.03394	&	10	&	3	&	1	&	0.56 	&	 3.88 	&	0.039	&	5.66 	&	3.24 	\\
9491	&	145.301666	&	11.511487	&	0.02161	&	8	&	2	&	3	&	4.34 	&	 1.09 	&	0.057	&	2.82 	&	7.60 	\\
9500	&	155.776886	&	13.007605	&	0.03217	&	21	&	6	&	1	&	0.59 	&	 3.64 	&	0.071	&	9.21 	&	6.11 	\\
9505	&	160.049683	&	13.582075	&	0.03187	&	11	&	7	&	1	&	0.60 	&	 1.50 	&	0.208	&	6.85 	&	3.22 	\\
9530	&	161.107101	&	14.093637	&	0.03282	&	20	&	4	&	1	&	0.17 	&	 11.70 	&	0.046	&	10.30 	&	1.99 	\\
9626	&	132.551987	&	29.418653	&	0.02697	&	9	&	5	&	6	&	5.72 	&	 1.57 	&	0.168	&	5.13 	&	20.90 	\\
10587	&	243.114563	&	29.415983	&	0.03203	&	24	&	11	&	2	&	2.11 	&	 16.40 	&	0.042	&	6.84 	&	13.10 	\\
10707	&	169.872375	&	12.043807	&	0.03945	&	13	&	3	&	1	&	0.71 	&	 2.60 	&	0.063	&	6.40 	&	4.86 	\\
10716	&	180.606689	&	12.338239	&	0.04066	&	8	&	6	&	2	&	1.01 	&	 2.06 	&	0.074	&	5.60 	&	8.32 	\\
10762	&	202.31102	&	11.593844	&	0.02273	&	38	&	16	&	7	&	6.44 	&	 14.70 	&	0.063	&	14.80 	&	18.50 	\\
10788	&	200.406204	&	12.22548	&	0.03802	&	8	&	6	&	3	&	5.50 	&	 2.46 	&	0.131	&	3.65 	&	29.50 	\\
11261	&	224.061127	&	9.425362	&	0.02889	&	15	&	9	&	3	&	1.57 	&	 2.40 	&	0.075	&	8.25 	&	9.15 	\\
11267	&	228.952332	&	8.317826	&	0.03388	&	15	&	8	&	4	&	4.09 	&	 1.27 	&	0.133	&	9.09 	&	25.30 	\\
11270	&	229.818726	&	8.331722	&	0.03387	&	9	&	4	&	1	&	0.58 	&	 4.37 	&	0.175	&	7.19 	&	4.88 	\\
11280	&	235.114166	&	7.447909	&	0.03771	&	9	&	6	&	4	&	3.20 	&	 2.68 	&	0.083	&	2.38 	&	24.40 	\\
11328	&	228.947556	&	9.012357	&	0.03344	&	14	&	11	&	4	&	3.20 	&	 1.73 	&	0.185	&	11.00 	&	18.50 	\\
11339	&	235.86792	&	8.283604	&	0.041	&	27	&	15	&	2	&	2.36 	&	 6.49 	&	0.177	&	22.40 	&	15.80 	\\
11378	&	217.050323	&	11.439986	&	0.02706	&	23	&	11	&	5	&	2.66 	&	 8.28 	&	0.064	&	8.52 	&	13.30 	\\
11438	&	226.04303	&	10.034683	&	0.03799	&	9	&	6	&	3	&	1.70 	&	 3.30 	&	0.073	&	5.50 	&	12.90 	\\
11464	&	221.710175	&	11.56801	&	0.02957	&	18	&	6	&	1	&	0.60 	&	 4.57 	&	0.037	&	12.20 	&	4.82 	\\
11473	&	227.018005	&	10.309496	&	0.0352	&	8	&	4	&	1	&	1.05 	&	 1.21 	&	0.072	&	5.49 	&	7.33 	\\
11705	&	204.111176	&	6.438471	&	0.02309	&	10	&	8	&	7	&	3.75 	&	 4.30 	&	0.088	&	2.80 	&	14.00 	\\
11726	&	203.263153	&	7.252505	&	0.02344	&	13	&	8	&	3	&	0.96 	&	 6.22 	&	0.039	&	5.37 	&	7.33 	\\
11985	&	219.562546	&	9.351056	&	0.03061	&	9	&	5	&	3	&	1.93 	&	 2.65 	&	0.065	&	7.53 	&	9.77 	\\
12133	&	233.813736	&	27.306461	&	0.03253	&	24	&	11	&	4	&	1.83 	&	 11.90 	&	0.049	&	12.30 	&	15.50 	\\
12281	&	235.330475	&	28.170858	&	0.03237	&	26	&	11	&	9	&	4.72 	&	 5.32 	&	0.076	&	16.10 	&	31.40 	\\
12418	&	241.956696	&	25.473261	&	0.04127	&	11	&	4	&	2	&	0.97 	&	 1.39 	&	0.053	&	8.14 	&	9.58 	\\
12512	&	226.913727	&	6.967523	&	0.03077	&	8	&	5	&	3	&	2.90 	&	 1.40 	&	0.11	&	4.28 	&	14.40 	\\
12515	&	229.197678	&	7.050813	&	0.03602	&	62	&	12	&	1	&	1.74 	&	 30.00 	&	0.043	&	36.00 	&	7.30 	\\
12524	&	203.16127	&	9.802313	&	0.02354	&	11	&	6	&	7	&	2.83 	&	 6.40 	&	0.094	&	4.73 	&	14.60 	\\
12530	&	212.721451	&	8.842516	&	0.02376	&	16	&	5	&	10	&	5.56 	&	 2.60 	&	0.178	&	6.80 	&	20.40 	\\
12534	&	220.137329	&	8.357473	&	0.03029	&	8	&	5	&	3	&	4.24 	&	 3.45 	&	0.067	&	1.75 	&	11.00 	\\
12537	&	223.078217	&	7.9186	&	0.03539	&	15	&	8	&	3	&	1.69 	&	4.92 	 &	0.079	&	9.06 	&	13.30 	\\
12577	&	210.567734	&	9.222879	&	0.02046	&	8	&	5	&	5	&	2.81 	&	 1.21 	&	0.127	&	2.45 	&	7.20 	\\
12588	&	220.664902	&	8.677205	&	0.03382	&	19	&	9	&	4	&	3.86 	&	 4.67 	&	0.073	&	20.90 	&	21.40 	\\
12598	&	232.917923	&	7.277811	&	0.03379	&	13	&	6	&	3	&	1.91 	&	 2.64 	&	0.069	&	6.27 	&	14.10 	\\
12711	&	234.154953	&	5.809239	&	0.03935	&	8	&	6	&	3	&	2.27 	&	 3.06 	&	0.111	&	8.83 	&	17.80 	\\
13585	&	201.159729	&	13.930791	&	0.02313	&	24	&	16	&	11	&	5.02 	&	 8.59 	&	0.077	&	9.79 	&	21.30 	\\
13622	&	209.656448	&	13.523097	&	0.03926	&	13	&	6	&	1	&	0.55 	&	 1.32 	&	0.065	&	6.09 	&	4.62 	\\
13680	&	208.240234	&	14.56538	&	0.04067	&	18	&	9	&	1	&	1.00 	&	 2.57 	&	0.097	&	20.80 	&	5.93 	\\
13702	&	208.900848	&	14.670437	&	0.04069	&	13	&	7	&	1	&	1.51 	&	 2.79 	&	0.124	&	10.20 	&	8.17 	\\
13870	&	221.791779	&	13.614258	&	0.0298	&	25	&	8	&	1	&	0.33 	&	 8.05 	&	0.033	&	14.00 	&	2.74 	\\
13876	&	226.666138	&	12.730574	&	0.02192	&	8	&	5	&	5	&	2.95 	&	 1.56 	&	0.042	&	2.59 	&	8.51 	\\
15023	&	142.037308	&	29.966383	&	0.02637	&	22	&	11	&	4	&	1.86 	&	 3.97 	&	0.084	&	21.30 	&	10.60 	\\
15033	&	129.532425	&	25.059904	&	0.0283	&	18	&	6	&	3	&	1.18 	&	 9.25 	&	0.051	&	7.34 	&	10.50 	\\
15056	&	138.513535	&	30.028185	&	0.02232	&	11	&	8	&	3	&	1.50 	&	 2.34 	&	0.047	&	6.29 	&	7.73 	\\
15831	&	139.646729	&	27.81535	&	0.0269	&	12	&	10	&	4	&	2.32 	&	 14.00 	&	0.061	&	3.99 	&	10.50 	\\
16375	&	230.110504	&	25.748169	&	0.03348	&	26	&	7	&	2	&	0.63 	&	 11.60 	&	0.049	&	11.10 	&	6.06 	\\
16770	&	218.138138	&	28.454185	&	0.03119	&	11	&	7	&	6	&	4.39 	&	 3.02 	&	0.133	&	2.82 	&	22.00 	\\
16825	&	205.954681	&	30.04793	&	0.04003	&	18	&	8	&	3	&	2.77 	&	 3.33 	&	0.081	&	13.60 	&	17.90 	\\
16987	&	164.100845	&	29.687904	&	0.0346	&	8	&	7	&	1	&	0.42 	&	 1.19 	&	0.164	&	7.32 	&	3.25 	\\
16990	&	166.251236	&	29.895872	&	0.02951	&	9	&	8	&	4	&	1.87 	&	 2.81 	&	0.066	&	4.52 	&	10.20 	\\

\hline											
\end{tabular}											
\end{table*}											
											
\addtocounter{table}{-1}											
\begin{table*}											
\centering											
\caption{The resulting values for the SDSS sample. (continued from previous page)}											
\begin{tabular}{lllllllllllll}											
\hline											
groupID &ra &dec &z & Nm &$N_{\rm spiral}$ &$N_{\rm HI}$ &$M_{\rm HI}$ &$M_{\rm v}$ &$t_{\rm c}H_{\rm 0}$ &$M_{\rm HI,color}$ & $M_{\rm HI,error}$ \\
        &   &    &  &    &                 &             &$\rm 10^{10} M_{\odot}$ &$\rm 10^{13} M_{\odot}$ &              &        $\rm 10^{9} M_{\odot}$     &$\rm 10^{8} M_{\odot}$  \\
\hline												
17428	&	196.622681	&	28.652262	&	0.02567	&	20	&	7	&	3	&	0.97 	&	 13.10 	&	0.122	&	5.30 	&	3.86 	\\
17438	&	204.825241	&	27.71818	&	0.03623	&	12	&	6	&	3	&	2.21 	&	 3.12 	&	0.092	&	7.00 	&	16.00 	\\
17450	&	229.861816	&	20.81197	&	0.03993	&	54	&	23	&	4	&	3.88 	&	 14.30 	&	0.092	&	32.10 	&	28.40 	\\
17454	&	231.037643	&	20.715727	&	0.03995	&	56	&	22	&	4	&	3.88 	&	 13.60 	&	0.101	&	22.30 	&	21.60 	\\
17459	&	194.460205	&	28.984585	&	0.02593	&	10	&	5	&	1	&	0.14 	&	 2.18 	&	0.047	&	8.19 	&	0.86 	\\
17470	&	213.153107	&	26.746155	&	0.03532	&	24	&	16	&	4	&	3.50 	&	 5.73 	&	0.154	&	18.50 	&	22.80 	\\
17506	&	216.144028	&	26.43553	&	0.03673	&	24	&	11	&	4	&	7.34 	&	 10.90 	&	0.087	&	10.30 	&	25.40 	\\
17535	&	242.944565	&	16.920317	&	0.03421	&	8	&	4	&	4	&	3.33 	&	 3.59 	&	0.066	&	3.88 	&	18.20 	\\
17587	&	208.999329	&	28.574835	&	0.03508	&	12	&	4	&	4	&	2.58 	&	 2.63 	&	0.075	&	4.70 	&	19.60 	\\
17785	&	208.208267	&	25.052908	&	0.02981	&	50	&	19	&	19	&	11.30 	&	 6.39 	&	0.172	&	22.50 	&	66.20 	\\
17810	&	241.82016	&	14.961396	&	0.03753	&	8	&	6	&	2	&	2.59 	&	 3.01 	&	0.077	&	9.72 	&	14.60 	\\
17813	&	242.940155	&	14.116739	&	0.03158	&	20	&	12	&	4	&	2.52 	&	 7.43 	&	0.095	&	16.50 	&	25.00 	\\
17839	&	239.485428	&	16.300982	&	0.03695	&	31	&	17	&	8	&	6.25 	&	 12.50 	&	0.036	&	23.50 	&	44.40 	\\
17843	&	241.614563	&	15.725222	&	0.03941	&	54	&	18	&	2	&	1.42 	&	 26.50 	&	0.048	&	35.00 	&	9.83 	\\
17866	&	208.996063	&	25.237631	&	0.03682	&	10	&	4	&	2	&	1.75 	&	 1.66 	&	0.075	&	5.95 	&	15.30 	\\
17951	&	245.345215	&	14.627812	&	0.02937	&	9	&	6	&	2	&	1.53 	&	 2.20 	&	0.069	&	5.35 	&	9.45 	\\
18257	&	239.609665	&	14.202219	&	0.0341	&	10	&	7	&	1	&	1.45 	&	 8.90 	&	0.092	&	10.50 	&	5.56 	\\
18297	&	241.863403	&	14.072877	&	0.03441	&	12	&	5	&	1	&	0.44 	&	 2.46 	&	0.118	&	10.50 	&	3.15 	\\
18513	&	134.122711	&	20.201704	&	0.03133	&	15	&	10	&	1	&	0.53 	&	 3.29 	&	0.116	&	18.20 	&	3.16 	\\
18625	&	137.515656	&	19.563103	&	0.02884	&	20	&	12	&	4	&	3.17 	&	 4.91 	&	0.079	&	16.00 	&	18.00 	\\
18627	&	139.505081	&	20.082504	&	0.02904	&	35	&	19	&	6	&	5.49 	&	 12.60 	&	0.097	&	21.00 	&	19.60 	\\
18629	&	155.366745	&	25.580858	&	0.02086	&	11	&	7	&	2	&	0.57 	&	 1.83 	&	0.125	&	2.24 	&	3.14 	\\
18633	&	162.123428	&	26.449635	&	0.02121	&	8	&	7	&	7	&	5.85 	&	 3.58 	&	0.08	&	0.44 	&	13.40 	\\
18655	&	185.056168	&	28.451227	&	0.02694	&	31	&	16	&	8	&	4.73 	&	 16.90 	&	0.059	&	11.00 	&	18.90 	\\
18728	&	167.791702	&	28.690138	&	0.02891	&	9	&	3	&	1	&	0.63 	&	 1.24 	&	0.049	&	5.97 	&	2.35 	\\
19077	&	146.832458	&	21.938108	&	0.02515	&	12	&	7	&	3	&	2.35 	&	 1.43 	&	0.117	&	5.92 	&	8.10 	\\
19146	&	175.772552	&	26.342564	&	0.03081	&	13	&	6	&	3	&	1.23 	&	 1.72 	&	0.088	&	8.27 	&	8.14 	\\
19153	&	188.938171	&	26.719774	&	0.02383	&	24	&	5	&	5	&	1.62 	&	 26.00 	&	0.027	&	10.30 	&	8.43 	\\
19166	&	167.841324	&	26.069723	&	0.03962	&	22	&	15	&	6	&	5.75 	&	 9.82 	&	0.095	&	15.60 	&	35.40 	\\
19195	&	193.259766	&	27.138609	&	0.02537	&	10	&	4	&	4	&	1.33 	&	 14.80 	&	0.085	&	3.34 	&	11.40 	\\
19206	&	169.401978	&	26.676378	&	0.02716	&	8	&	5	&	3	&	2.83 	&	 3.43 	&	0.067	&	3.52 	&	12.70 	\\
19207	&	171.896118	&	27.026323	&	0.03344	&	8	&	7	&	3	&	1.84 	&	 1.86 	&	0.212	&	5.80 	&	17.20 	\\
19222	&	193.218887	&	27.089399	&	0.02113	&	8	&	5	&	1	&	0.15 	&	 3.08 	&	0.068	&	4.11 	&	1.71 	\\
19245	&	167.262909	&	26.547382	&	0.03808	&	17	&	15	&	6	&	5.32 	&	 4.75 	&	0.091	&	13.20 	&	30.30 	\\
19258	&	181.580505	&	28.11812	&	0.02802	&	35	&	10	&	3	&	0.85 	&	 30.60 	&	0.044	&	15.60 	&	6.31 	\\
19301	&	179.667572	&	28.237883	&	0.02759	&	11	&	6	&	1	&	2.82 	&	 4.03 	&	0.101	&	3.72 	&	3.50 	\\
19607	&	183.889755	&	23.987774	&	0.02244	&	19	&	10	&	5	&	2.60 	&	 7.72 	&	0.045	&	6.46 	&	9.18 	\\
19631	&	182.067291	&	25.323114	&	0.02242	&	32	&	10	&	2	&	0.48 	&	 7.22 	&	0.088	&	12.70 	&	3.67 	\\
19633	&	184.897842	&	25.436333	&	0.02291	&	13	&	9	&	6	&	2.20 	&	 4.22 	&	0.113	&	3.58 	&	11.30 	\\
19655	&	173.281143	&	25.118263	&	0.03337	&	9	&	6	&	6	&	4.28 	&	 4.42 	&	0.057	&	2.42 	&	26.90 	\\
19709	&	139.018967	&	17.620777	&	0.03102	&	8	&	4	&	1	&	0.33 	&	 2.51 	&	0.056	&	4.69 	&	3.91 	\\
19710	&	139.187943	&	17.268932	&	0.02837	&	32	&	15	&	3	&	1.07 	&	 7.52 	&	0.159	&	20.30 	&	8.69 	\\
19725	&	121.261459	&	10.62543	&	0.03445	&	11	&	7	&	6	&	7.02 	&	 3.84 	&	0.085	&	6.06 	&	40.40 	\\
19767	&	156.791946	&	21.746721	&	0.0415	&	14	&	7	&	1	&	1.32 	&	 1.00 	&	0.14	&	13.40 	&	7.91 	\\
19826	&	164.541595	&	24.242392	&	0.02119	&	9	&	2	&	3	&	0.62 	&	 2.90 	&	0.037	&	3.03 	&	5.20 	\\
19892	&	171.01944	&	24.125832	&	0.02481	&	20	&	15	&	5	&	2.54 	&	 17.80 	&	0.056	&	8.74 	&	13.40 	\\
19958	&	121.009537	&	9.955961	&	0.03389	&	14	&	8	&	1	&	1.15 	&	 2.60 	&	0.064	&	12.20 	&	5.82 	\\
20006	&	177.568878	&	21.269369	&	0.02605	&	10	&	5	&	6	&	1.77 	&	 1.06 	&	0.124	&	3.85 	&	14.00 	\\
\hline											
\end{tabular}											
\end{table*}											
											
\addtocounter{table}{-1}											
\begin{table*}											
\centering											
\caption{The resulting values for the SDSS sample. (continued from previous page)}											
\begin{tabular}{lllllllllllll}											
\hline											
groupID &ra &dec &z & Nm &$N_{\rm spiral}$ &$N_{\rm HI}$ &$M_{\rm HI}$ &$M_{\rm v}$ &$t_{\rm c}H_{\rm 0}$ &$M_{\rm HI,color}$ & $M_{\rm HI,error}$ \\
        &   &    &  &    &                 &             &$\rm 10^{10} M_{\odot}$ &$\rm 10^{13} M_{\odot}$ &              &    $\rm 10^{9} M_{\odot}$   &$\rm 10^{8} M_{\odot}$  \\
\hline												
20006	&	177.568878	&	21.269369	&	0.026	&	10	&	5	&	6	&	1.77	 &	 0.81	&	0.150	&	3.85	&	14	\\
20008	&	178.070297	&	20.775753	&	0.022	&	40	&	23	&	20	&	6.23	 &	 9.3	&	0.097	&	15.8	&	32.6	\\
20015	&	183.291992	&	21.631865	&	0.024	&	9	&	2	&	1	&	0.68	 &	 3.86	&	0.036	&	3.97	&	2.15	\\
20024	&	222.336075	&	17.333218	&	0.038	&	19	&	16	&	6	&	4.48	 &	 6.01	&	0.117	&	14.9	&	38.1	\\
20036	&	144.31813	&	17.022217	&	0.028	&	11	&	6	&	1	&	0.69	 &	 4.8	&	0.103	&	5.27	&	2.79	\\
20153	&	222.130249	&	18.291761	&	0.039	&	21	&	11	&	1	&	3.09	 &	 2.29	&	0.077	&	13.8	&	9.33	\\
20157	&	225.826508	&	17.168579	&	0.039	&	9	&	7	&	2	&	1.45	 &	 1.41	&	0.118	&	8.4	&	10.6	\\
20211	&	208.037155	&	21.605583	&	0.028	&	28	&	16	&	5	&	2.33	 &	 11.4	&	0.082	&	15.1	&	15.6	\\
20304	&	135.91922	&	13.621006	&	0.029	&	21	&	12	&	6	&	2.14	 &	 6.3	&	0.100	&	10.4	&	16.3	\\
20436	&	240.615356	&	12.548163	&	0.035	&	13	&	7	&	3	&	2.63	 &	 3.75	&	0.138	&	10.5	&	17.9	\\
20728	&	176.801102	&	18.731853	&	0.037	&	8	&	4	&	1	&	1.05	 &	 1.83	&	0.118	&	9.31	&	6.32	\\
20749	&	181.065323	&	20.481783	&	0.024	&	64	&	30	&	17	&	7.76	 &	 17.2	&	0.072	&	34.8	&	35.6	\\
20831	&	205.821304	&	17.955034	&	0.027	&	13	&	4	&	4	&	1.33	 &	 1.95	&	0.052	&	7.07	&	11.4	\\
20856	&	237.042297	&	11.783937	&	0.035	&	14	&	7	&	3	&	2.85	 &	 3.67	&	0.088	&	6.16	&	17	\\
20924	&	238.412888	&	12.088135	&	0.035	&	11	&	10	&	6	&	9.22	 &	 1.17	&	0.176	&	10	&	35.4	\\
21006	&	154.255188	&	16.947313	&	0.028	&	8	&	6	&	2	&	1.53	 &	 2.85	&	0.124	&	6.14	&	6.12	\\
21012	&	164.634491	&	18.140448	&	0.031	&	11	&	7	&	6	&	4.62	 &	 4.08	&	0.129	&	5.34	&	22.1	\\
21045	&	161.018997	&	18.49087	&	0.031	&	8	&	6	&	5	&	4.82	 &	 0.89	&	0.074	&	1.34	&	26.9	\\
21050	&	165.82312	&	18.841419	&	0.032	&	16	&	11	&	6	&	3.27	 &	 6.27	&	0.127	&	6.96	&	23.9	\\
21308	&	212.117477	&	14.969675	&	0.026	&	12	&	9	&	8	&	5.73	 &	 1.22	&	0.316	&	4.45	&	22	\\
21635	&	162.619339	&	16.014072	&	0.021	&	9	&	3	&	2	&	1.77	 &	 3.81	&	0.053	&	4.13	&	3.92	\\
21680	&	164.388306	&	17.183245	&	0.031	&	13	&	6	&	4	&	2.97	 &	 5.8	&	0.074	&	13.8	&	15.7	\\
22080	&	152.140274	&	14.882695	&	0.030	&	9	&	7	&	2	&	1.24	 &	 1.07	&	0.185	&	9.7	&	7.89	\\
\hline											
\end{tabular}%
\end{table*}




\label{lastpage}
\end{document}